\documentclass[11pt]{article}
\usepackage{graphicx}
\usepackage{epsfig, amsmath, amssymb}
\usepackage{epstopdf}

\title{Thorat Finding Algorithms based on Throat Types}%
\begin{document}
\begin{center}
\LARGE{\bf{Throat Finding Algorithms based on Throat Types}}
\end{center}

\begin{center}
Kyung-Taek Jun \\  Department of Applied Mathematics, Stony Brook University, NY \\ ktfriends@gmail.com
\end{center}
\vspace{1in}

\section{Introduction}
Understanding the precise structure of pore space is integral to fully understanding flow through porous media. This understanding is essential in research on single-phase incompressible flow in situations where another dominant flow factor is the fluid-wall surface tension \cite{lind1}. Determining the relationship of pore structure and bulk flow properties is complicated since the structure of pore space is so very random and awkward because of the challenge of precisely measuring pore space. 

A throat is the smallest corss-section area that corresponds a branch-branch medial axis path.  For non-crossed throats, their outer perimeter voxels have to exist on the boundary grain voxels. The 3DMA-Rock software package \cite{lind2} has three major throat-finding algorithms; (1) the wedge-based algorithm \cite{shin},(2) the Dijkstra-based shortest length algorithm \cite{lind1}, and (3) the planar dilation algorithm \cite{jw}. The wedge-based algorithm yields acceptable measures in only low porosity samples. It uses wedges to find the nearest boundary grain voxels and connects voxels in a way that allows the determination of border voxels which allow the connection of each wedge. For high porosity samples (over 30$\%$), It is very diffcult to find the connedcted throat-perimeter paths on the boundary grain voxels. The primary deficiency of the Dijkstra-based shortest length algorithm is that the shortest path perimeter does not reveal the smallest cross area (Figure 1). The main deficiency of the planar dilation algorithm is in finding approximated throats. In general, the throat shape is a non-planar type. The throat determined by the employment of the planar dilation algorithm is capable of spreading pore space without any restriction.

The goal in this paper is to find the accurate throat area (Figure 2 right panel) and calculate the circumference based on known mathematical formulae. Here, I present new throat-finding algorithms using the concepts in vector space and spherical coordinate system (decribed in section 2). In order to find an accurate throat, I need to classify throat shapes into three main types; (1) the simply-connected planar type, (2) the simply-connected non-planar type, and (3) the non-simply-connected non-planar type. For each thorat shape, I construct five algorithms; two of them for the simply-connected planar type, two of them for the simply-connected non-planar type, and one of them for the non-simply-connected non-planar type. The first algorithm is designed to reduce the total calculation time. It indicates the simply-connected planar throat perimeter, and only uses the visible path (the member voxels of the path are visible from the MA path, invisible path is not) from the pertinent MA path. Usually the perimeter (the solution of the first algorithm) is not a throat, but we can reduce the total calculated CPU time (involving 4 other of our algorithms) using this possible candidate perimeter. The second algorithm indicates the simply-connected planar throat perimeter. This second algorithm uses visible paths and invisible paths (from the medial axis voxel) that together make an enclosed loop. All 26-connected perimeter voxel paths are on the plane. The third algorithm is appropriate for determining the simply-connected non-planar throat perimeter. The perimeter voxels, in cases where we employ this third algorithm, consist of visible voxels and invisible voxels from the pertinent medial axis path. The visible voxels are on the plane, and the third algorithm connects discontinuous voxels using non-planar voxels (generally, the voxel set is not on the same plane of visible voxel set). The fourth algorithm is designed to determine the largely undulate throat perimeter (See Fig. 2). This algorithm makes four vertical and horizontal wedges in each neighboring direction (based on the normal plane to the medial axis path). For each wedge, the third algorithm finds the nearest voxels to the MA voxel, and connects eight perimeter voxels using Dijkstra’s algorithm \cite{dij}. The fifth algorithm is designed to determine non-simply non-planar throat. Also, this algorithm is useful when a small number of visible voxels diverge from the perimeter. After finding the visible continuous voxel set, the fifth algorithm tries to connect them. When this algorithm fails to connect voxels, we delete the voxels from the member of the candidate perimeter voxel set; we keep the path with the shortest distance from the MA path when there are two or more paths. To deal with discontinuous voxels in the perimeter voxel set, this algorithm uses Dijkstra’s algorithm. When Dijkstra’s algorithm finds several paths, our algorithm uses the innermost path from the medial axis voxel that results in affording the smallest throat area. These algorithms can detect more than 98$\%$ throats over higher than 29 $\%$ porosity samples.

\section{Throat-Finding Algorithms}
\subsection{Simply Connected Planar Throat-Finding algorithms}

The first algorithm is to find throat on a plane, called planar throat type. In this algorithm, both medial axis modiification \cite{lind2} and medial axis extraction \cite{lee}, which computed within the void space of a segmented \cite{oh} 3D XCMT image, are used. For constructing the first throat-finding algorithm, I need to check all possible perimeters of a throat from the first  to the last voxel in the path of the medial axis since the area of throat does not show continuously change. The area of a throat can be varied by the position and direction of the medial axis voxels in three dimensional space. Here, I present two simply connected planar throat-finding algorithms. The first algorithm can be used when the boundary grain voxel set is visible from the medial axis path. The second algorithm is used when some voxels in the boundary grain voxel set is not visible from the medial axis path.

\subsubsection{The first Simply Connected Planar Throat-Finding algorithm}

The first algorithm is the following steps. Here, I define \newline
(1) $\nu_{k}$ : $k$-th voxel of the medial axis path, \newline
(2) $\overrightarrow{n_{k}}$ :  unit tangent vector of the medial axis path,  \newline
(3) $\overrightarrow{n_{k,i.j}}$ : unit directional vector where $i$ is a degree of polar angle from  $\overrightarrow{n_{k}}$  and $j$ is a degree of azimuth angle in spherical coordinate system  \newline
(4) $P_{k,i,j}$ : a plane that has the normal vector  $\overrightarrow{n_{k,i.j}}$ at the center of $\nu_{k}$ \newline
(5) $\overrightarrow{u_{k,i,j,\theta}}$  : unit directional vector on the plane     $P_{k,i,j}$

Step 1 : Compute the tangent vector $\overrightarrow{n_{k}}$ at the center of $\nu_{k}$ voxel in the path

Step 2 : Make unit directional vector set of $\overrightarrow{n_{k,i,j}}$ with zenith vector  $\overrightarrow{n_{k}}$ 

Step 3 : Make a plane $P_{k.i,j}$ with a point at the center of $\nu_{k}$ and a normal verctor $\overrightarrow{n_{k,i,j}}$

Step 4 : Make 360 of unit directional vector set $\overrightarrow{u_{k,i,j,\theta}}$ on  $P_{k,i,j}$

Step 5 : Find the boundary grain voxel set using $\overrightarrow{u_{k,i,j,\theta}}$ (Finding method will be discussed in below)

Step 6 :  Check whether the connectivity of the boundary grain voxel set from Step 5 has 26 connectivity or not

Step 7 : If the connectivity of the boundary grain voxel set shows 26 connectivity then extend it to 6 connected closed loop voxel set. Otherwise, I use different throat-finding algorithms (these algorithms will be discussed in section 2.1.2 and section 2.2.1)

Step 8 : Calculate the area of the closed loop voxel set using 26 connected perimeter voxel set from Step 6.

Step 9 : Repeat from Step 1 to 7 for each $\nu_{k}$  to get local minimum throat

The main idea of this algorithm is to use the visible boundary grain voxels from the medial axis voxel on a given plane. All voxels originate as specific points within each coordinate. For the medial axis voxel $\nu{_k}$, $k = 1, 2, 3, \cdots, n$, where $n$ is the last voxel number in the medial axis path, we choose 5 consecutive medial axis voxels (the four proximal – at right and left, two at each side) to find the zenith direction $n_{k}$, which is the solution of the least square problem of the 5 involved voxels. If the medial voxel position is located within the two outermost voxels on each side, I choose the 5 outermost in the medial axis path that includes the medial axis voxel (Figure 4 B). If the total number of voxels in the path is less than 5, I use them to calculate the directional vector solution of the least square problem. Using the center of $\nu_{k}$ = $(\nu_{x_{k} },\nu_{y_{k}},\nu_{z_{k}})$ and $\overrightarrow{n_{k}}$ =  $<n_{x_{k}},n_{y_{k}},n_{z_{k}}>$,  I can make the sequence $\overrightarrow{n_{k,i,j}}$ (Figure 4 C), defined by the polar and azimuth angles, $i$ and $j$, $i$=1,2,$\cdots$, 45, $j$ = 1,2,$\cdots$, 45, including $n_{k}$. The center of $\nu_{k}$ and its normal vector $\overrightarrow{n_{k,i,j}}$ create the plane.  I now need to make two unit vector sets: $\{u_{k,i,j,\theta}\}$ and $\{s_{k,i,j,\theta}\}$ where $\theta$ = $0,1, \cdots, 359$ (Figure 4 D). The ray set $\{u_{k,i,j,\theta}\}$ on the plane $P_{k,i,j}$  is normalized by the orthogonal projection of the unit vector set $\{s_{k,i,j,\theta}\}$. The unit ray set $\{s_{k,i,j,\theta}\}$ has an angle difference of $1^{o}$, and is on the coordinate plane that consists of the two coordinate axes that have two small absolute components of $\overrightarrow{n_{k,i,j}}$. The set  $\{s_{k,i,j,\theta}\}$ uses the formula of one of the three different sets:
\begin{eqnarray*}
\{(cos(\theta^{o}), sin(\theta^{o}),0)\}, \{(cos(\theta^{o}), 0, sin(\theta^{o}))\}, \{(0, cos(\theta^{o}), sin(\theta^{o}))\}
\end{eqnarray*}
where $\theta$ = $0,1,\cdots, 359$.

I choose the set $\{(cos(\theta^{o}), sin(\theta^{o}),0)\}$  when $|n_{z_{k,i,j}} |\ge |n_{x_{k,i,j}}|$  and  $|n_{y_{k,i,j}}|$. The two others are selected similarly. The directional vector $\overrightarrow{u_{k,i,j,\theta}}$ = $<u_{x_{k,i,j,\theta}},u_{y_{k,i,j,\theta}},u_{z_{k,i,j,\theta}}>$ starts from the center of $\nu_{k}$ and spreads until it touches the respective boundary grain voxel (Figure 5). Each ray yields three coefficient sets: $\{c_{x_{\alpha_{1}}}\}$, $\{c_{y_{\alpha_{1}}}\}$, and $\{c_{z_{\alpha_{1}}}\}$, $\alpha_{i}$ = $1,2,3$, $\cdots$ and $i$ = 1,2,3 that satisfy $|u_{\beta_{k,i,j,\theta}}|$$\cdot c_{\beta_{\alpha_{t}}}$ = $t-0.5$, $t=1,2,\cdots$ and $\beta = x, y, z$, when $u_{\beta_{k,i,j,\theta}}$ is not zero.If $u_{\beta_{k,i,j,\theta}}$ is zero, then the ray doesn’t go in the β coordinate direction. Those coefficients represent the boundary points that touch the voxels on the ray. Using the mixed set of $\{c_{x_{\alpha_{1}}}\}$, $\{c_{y_{\alpha_{2}}}\}$, and $\{c_{z_{\alpha_{3}}}\}$ with increasing order, we can determine the voxel order that the pertinent ray touches. The order of the new set, at each point, indicates the direction of the voxel movement from $\nu_{k}$ If two coefficients among the mixed set are the same, then the ray touches the edge of the next voxel, with the related coordinates indicating the coefficients from the current position. If three coefficients are the same, then the ray touches the vertex of the next voxel, with the related coordinates indicating the coefficients from the current position. When the ray touches an edge or a vertex, we only check the next position to see whether the voxel is grain or void on the ray, because the void space has 26- connectivity. When we construct a throat barrier with 6-connectivity using a ray, we add all touching void voxels. If a ray touches an edge and all contiguous voxels are void, then the three void voxels will be counted among the throat barrier voxels. If a ray touches a vertex and all contiguous voxels are void, then the seven void voxels will be counted among the throat barrier voxels. The ray set $\overrightarrow{u_{k,i,j,\theta}}$ is used to search for the boundary grain voxels that are the candidates for the perimeter voxel set on $P_{k,i,j}$ (Figure 4 E). Usually finded grain voxel set has 26-connectivity, so we change the connectivity to the grain’s (6-connectivity). The set is comprised of blue voxels to the 6-connected set by adding the neighboring diagonally positioned boundary grain voxels on the plane. Sometimes an added voxel  is not on $P_{k,i,j}$. The candidate perimeter voxel set (Figure 4 F). The set is comprised of green Voxels) consists of piecewise 6-connected parts. When the 6-connected perimeter voxel set makes a closed loop, we calculate the area with the 26-connected voxel set. After we find all possible perimeters for all of the plane   $P_{k,i,j}$ , we compare all of them to find a possible candidate throat perimeter.

\subsubsection{The Second Simply Connected Planar Throat-Finding algorithm}

The second planar throat-finding algorithm follows the same process as the first, which constructs the 6-connected perimeter voxel set. This algorithm calculates the triangular area using the centers of the finded 26-connected boundary grain voxel set and $\nu_{k}$. This algorithm allows the progression to the next step, when the summation of the triangular areas is smaller than the minimum area of the first planar throat-finding algorithm (Figure 6 B). This algorithm finds invisible boundary grain voxels (for the unseen, distended perimeter) using Dijkstra’s algorithm. The candidate perimeter voxel set  (Figure The set is comprised of green Voxels) consists of piecewise 6-connected parts. For each discontinuous path (Figure 6 C Two orange voxels), we use Dijkstra’s algorithm to connect the end path voxels to the boundary grain voxels. This connection can be understood conceptually in the following way: The boundary grain voxels are on the plane  $P_{k,i,j}$. The basic criteria used to determine that a grain voxel is on  $P_{k,i,j}$ is $P_{k,i,j}$$(\nu_{x_{k} },\nu_{y_{k}},\nu_{z_{k}})$* $P_{k,i,j}$$(\nu_{x_{k} }+\epsilon_{x},\nu_{y_{k}}+\epsilon_{y},\nu_{z_{k}}+\epsilon_{z})$ $\le 0$, where $\epsilon_{x}$, $\epsilon_{y}$, and $\epsilon_{x}$ are 0 or 1, and at least one 1 is included. The region of interest concerning Dijkstra’s algorithm is conceptually within a cube of variable side length. The side length changes from $l_{c}+4$ to $l_{c}+8$, where $l_{c}$ is the side length of a minimum inclusion cube sharing the same center (Figure 6 C $l_{c}$ is the side length of the rose square). For each discontinuity, if there is no voxel connection in the cube, we assume it is a non-perimeter and we try to apply other algorithms. When we have found the perimeter voxel set that has 6-conectivity making a single and efficient encirclement (Figure 6 D Green voxels), we calculate the area with the 26-connected perimeter voxel set to find the minimum surface area using 3DMA. The 26-connected perimeter voxel set is selected within the 6-connected perimeter voxel set. After we find all possible perimeters for all of the plane  $P_{k,i,j}$, we compare all of them to find a possible candidate throat perimeter. Our algorithm has only one rounding number, and applies Dijkstra’s algorithm to the small discontinuous consecutive voxels.

\subsection{Simply Connected Non-Planar Throat-Finding algorithms}

I made two throat-finding algorithms for the undulating throat type. The first undulating throat-finding algorithm is designed to find simply connected throats. Sometimes, this algorithm cannot find the smallest throat area, but it can indicate the throat perimeter when the other three throat-finding algorithms fail to find the perimeter.

\subsubsection{The first Simply Connected Non-Planar Throat-Finding algorithm}

The first non-planar throat-finding algorithm is the same as the planar throat-finding algorithm, except concerning the region of interest in applying Dijkstra’s algorithm in dealing with the nearest boundary grain voxel from medial axis voxel $\nu_{k}$ for each step. For this algorithm, we select as the possible candidate the perimeter of a (conceptual) cube where the side length is the minimum cube $l_{c}+4$ that envelops two voxels with the same center in 3D (Figure 7 A). Now, 3DMA calculates the area with the 26-connected perimeter (Figure 7 B), derived from the 6-connected perimeter. After we find all possible perimeters for all of the plane $P_{k,i,j}$, we compare all of them to find a possible candidate throat perimeter.

\subsubsection{The second Simply Connected Non-Planar Throat-Finding algorithm}

The second non-planar algorithm can be used to analyze largely undulate type throats. We allow that the maximum angle of the fluctuation of this algorithm is $\pm45^{o}$ from the base. Our assumption is that some perimeter voxels are situated nearest to an MA voxel. If the throat undulates a lot and satisfies our assumptions, then this algorithm can deduce its perimeter.
Before constructing the plane $P_{k,i,j}$, we use the same procedure as the plane $P_{k,i,j}$ of the planar algorithm. We draw the 91 unit ray set $\{u_{k,i,j,\theta}\}$ , θ=0,1,⋯,90, on the plane $P_{k,i,j}$, drawn from the projections and normalization of the 91 unit ray set $\{s_{k,i,j,\theta}\}$, $\theta$ = $0,1, \cdots, 90$, on the coordinate axis plane (Figure 8 A). The set $\{s_{k,i,j,\theta}\}$ is selected among the three different sets:
\begin{eqnarray*}
\{(cos(\theta^{o}), sin(\theta^{o}),0)\}, \{(cos(\theta^{o}), 0, sin(\theta^{o}))\}, \{(0, cos(\theta^{o}), sin(\theta^{o}))\}
\end{eqnarray*}
where $\theta$ = $0,1,\cdots, 90$.. This 90-degree angle makes the first wedge. The selected order of the set $\{s_{k,i,j,\theta}\}$ is the same order of the maximum component value of $\overrightarrow{n_{k,i,j}}$. The set $\{u_{k,i,j,\theta}\}$ on $P_{k,i,j}$ is generated by the orthogonal projection of $\{s_{k,i,j,\theta}\}$ onto the plane. Using the two related ray sets $\{u_{k,i,j,\theta}\}$  and $\{s_{k,i,j,\theta}\}$, we find the nearest grain boundary voxel $p_{1}$ and its angle ($\epsilon$ =  $\alpha_{1}$) from $\nu_{k}$. The voxel $p_{1}$ is on $P_{k,i,j}$ and in the first wedge with the $90^{o}$ angle. The next step is making three horizontal wedges on $P_{k,i,j}$ (Figure 8 B). Their center angles are  $\epsilon+90^{o}$, $\epsilon+180^{o}$, and $\epsilon+270^{o}$ and their respective wedge angles are each $60^{o}$. The three 61-unit ray sets are selected among $\{s_{k,i,j,\theta}\}$ (Figure 8 B). The selected angle sets are below:
\begin{eqnarray*}
\{\theta_{2} |\theta_{2} = \epsilon+60^{o},\epsilon+61^{o}, \cdots, \epsilon+120^{o}\},\\
\{\theta_{3} |\theta_{3} = \epsilon+150^{o},\epsilon+151^{o}, \cdots, \epsilon+210^{o}\}\\
\{\theta_{4}|\theta_{4}= \epsilon+240^{o},\epsilon+241^{o}, \cdots, \epsilon+300^{o}\}
\end{eqnarray*}

For each horizontal wedge, we find the three nearest boundary grain voxels $p_{3}$, $p_{5}$, $p_{7}$ and their angles $\alpha_{i}$ ($i$=2,3,4) from $\nu_{k}$ on $P_{k,i,j}$ using 61 rays $\{u_{k,i,j,\theta}\}$ (Figure 8 B). We add four vertical wedges with center angles {$\beta_{j}$},$j$=1,2,3,4, where the angles are bisected between each angle of the set {$\alpha_{i}$}, $i$ = 1,2,3,4. Each vertical wedge’s azimuth angle varies from $45^{o}$ to $135^{o}$ from the new zenith $\overrightarrow{n_{k,i,j}}$.   We make a 91-ray set in the vertical wedge, where each unit vector’s azimuth angle varies from $45^{o}$ to $135^{o}$ (Figure 8 C). The first vertical unit ray set $\{u_{k,i,j,\theta}\}$ is $\overrightarrow{n_{k,i,j}}$, where $i = \alpha_{1}$ and $j$ =$\varphi+45^{o},\varphi+46^{o}, \cdots, \varphi+135^{o}$. The other three vertical unit ray sets are chosen similarly.  Using four 91 ray sets, we find the four nearest boundary grain voxels $p_{2}$, $p_{4}$, $p_{6}$, and $p_{8}$ to $\nu_{k}$. There are eight candidate perimeter voxels. The final step is to connect those 8 voxels using Dijkstra’s algorithm in dealing with the nearest boundary grain voxel from the medial axis voxel $\nu_{k}$ for each step (Figure 8 D). If there is a complete 6-connected perimeter, 3DMA calculates the area with a 26-connected perimeter(Figure 8 D).

\subsection{Non-Simply Connected Non-Planar Throat-Finding algorithm}

The MA algorithm only preserves the topological structure of the given void space, so a throat’s type cannot be simply connected sometimes. For example, when there is a large peak extending from the bottom. If the path length is shorter than the peak, and the peak contains the projection of the path and its cluster voxels, then the throat has to be considered a non-simply connected type. This algorithm finds some other throat types, and is excellently suited to do so. Also, this algorithm can be used when perimeter parts indicate other pores’ boundaries. At this time, the wrong path parts can be nixed.
When the throat type is the same as the previous type, the fourth algorithm can find similar perimeters. For the non-simply-connected non-planar throat type, we use the same procedure to find the plane $P_{k,i,j}$ and the 6-connected candidate perimeter set as the second planar algorithm. The set { The 6-connected voxel set is comprised of blue voxels). The set consists of piecewise continuous paths (Figure 9 C). This algorithm starts with a continuous path among the set that constitutes the shortest voxel distance in the set from $\nu_{k}$. Also, the starting voxel can be selected as the longest path or the largest angle. The algorithm connects each piece using Dijkstra’s algorithm in dealing with the nearest boundary grain voxel from ma voxel ν for each step. When this, our forth algorithm, fails to connect two pieces (Figure 9 D), the algorithm deletes a part of the set (Figure 9 D brown voxels). By deleting the discontinuous path, we connect to the next path (Figure 9 E). This algorithm’s candidate perimeter consists of the compositions of the piecewise continuous parts of {$p_{i}$} ({$q_{i}$} is a visible perimeter voxel set, and {$p_{i}$} is 6-connected voxel set that comes from {$q_{i}$}) and their connection (Figure 9 F). Sometimes this algorithm deletes lots of continuous paths of {$p_{i}$}, so it requires checking the rounding number. The rounding number is defined as the sum of the angles of the projected perimeter voxels on $P_{k,i,j}$. The perimeter has to have one rounding number. Now, the algorithm changes the perimeter to 26-connetivity, and 3DMA calculates the outer area.
If the set {$o_{i}$} - i=1,2,⋯, the total number of the perimeter voxels - is of the outer perimeter voxels of this type, the subtraction of the inner grain voxels’ area is necessary to calculate the throat area. We make a unit ray set {$\overrightarrow{b_{i}}$}, i=1,2,⋯, the total number of the perimeter voxels. A unit ray $\overrightarrow{b_{i}}$ ̂ indicates the respective position of the perimeter voxel relative to the MA voxel. Each $\overrightarrow{b_{i}}$ ̂ spreads to the perimeter voxel $o_{i}$ from $\nu_{k}$. Our algorithm works when $\overrightarrow{b_{i}}$̂ touches the edge or vertex of voxels on the ray. When $\overrightarrow{b_{j}}$ ̂ and $\overrightarrow{b_{j+1}}$̂ go through the grain voxels (Figure 9 G), our algorithm saves the initial and final positions. We subtract the quadrilateral area from the outer area (Figure 9 H).

\section{Point Set algorithm}

The perimeter is used for the simulations of the drainage processes. The entry condition equation involves the area and perimeter. Digitized images are constituted of discrete values for each voxel. After segmentation, we only know from the data whether the voxel is grain or not. The segmented images consist of cubes, so we have to think about the difference between our calculations from the segmented images and the real length and area. Construction of the polygon (which is the boundary of the object in the CT image) by linear connection of the mid-points yields the relative error of its length for any random boundary. The reason for the relative error is that the mid-point connection contains only horizontal, vertical, and diagonal lines. A line integral needs the slope between two points $a_{n}$=($x_{n}$,$y_{n}$), $a_{n+1}$=($x_{n}$+$\Delta x$,$y_{n}$+$\Delta y$),  so that the calculation expresses the real boundary length. The mid-point calculation of the length in CT images may yield a relative error of more than 8$\%$. Thus we needed to find a new method to calculate length with a new point set in the CT images. We use the 6-connected perimeter voxel set to calculate the exact length, and we use the 26-connected perimeter voxel set to calculate the area in 3D space. The mathematical concepts of the point set consist of: i) Differentiability, ii) Implicit Function Theorem, iii) Line Integral. For non-differentiable parts of the perimeter, we follow the inner part of the perimeter. When we apply a line integral to a CT image, we require only one mid-point on each $\Delta x$ instead of multiple points. (The extant mid-point method involves several horizontal and vertical points on each $\Delta x$.) For each $\Delta x$, the length is a multiple of the voxel size and all y values are the same. The mid-point locates the adjacent face of the perimeter voxel, and the point used is the middle of $\Delta x$ within part of the perimeter voxel set. We only use the perimeter voxel part that faces the 6-connected barrier voxel set in 3D space. To easily understand our algorithm, we here give an example for a line integral in a CT image in Figure 13 A and B. 
We have to know which parts of the perimeter voxel set are related to the mathematical concepts of differentiability, line integral, and implicit function theorem. Using these math concepts, we select the point set on the boundary between perimeter voxels and throat barrier voxels, because the perimeter voxels are located outside of the throat. We will divide the perimeter voxel set into several parts to apply these three concepts.  The first step of the new algorithm distinguishes which perimeter parts are related to the differential and non-differential functions. We start with these. For the non-differential parts (U, T, and L shapes), we select points that identify the perimeter section shape. Those shapes consist of the combination of continuous horizontal and vertical voxels, and each part of the (respective) continuous vertical and horizontal part has at least 3 voxels.  The first point is on the inner-facing part of the initial voxel, in the non-differential part in 2D space. From the second point on, points are selected: at each right angle in the non-differential perimeter part. Let {an} be a perimeter array and {$p_{i}$} be a point set, and let {$a_{k_{n}}$} be a subset of {$a_{n}$}. U and T shaped voxels consist of horizontal-vertical-horizontal or vertical-horizontal-vertical voxel parts, and each part has the same x or y value in each of at least three voxels. To classify the U and T shaped voxels, a possible condition is $k_{2}$ – $k_{1}$ $\geq$ 2, $k_{3}$ – $k_{2}$ $\geq$ 1, and $k_{4}$ – $k_{3}$ $\geq$ 2. The point  $a_{k_{i}}$ (1 $\leq$  i $\leq$ 4) indicates vertex points (Figure 10 A and B). If the distances between the created points from each two sequential voxels are greater than or equal to 2 voxels, then the part satisfies the U and T shape conditions and four points are added into the point set. There are eight possible basic U and T shapes. The L shaped perimeter voxel part consists of horizontal-vertical or vertical-horizontal voxel parts, and each part has the same x or y value in each of at least three voxels. To classify the L shaped perimeter voxel part, one possible condition is $k_{2}$ – $k_{1}$ $\geq$ 2 and $k_{3}$ – $k_{2}$ $\geq$ 2. This process is similar to the process of finding the U and T shapes. Three points $a_{k_{1}}$, $a_{k_{2}}$, and $a_{k_{3}}$ are our vertex point. If the distances between the created points from each two sequential voxels are greater than or equal to 2 voxels, then the part satisfies the L shape condition and three points are added into the points set (Figure 10 C). The second step of the new algorithm distinguishes the perimeter part related to the implicit function theorem that will give us the C shape. The distinguishing condition of the C shape is that two edge voxels of the vertically altered perimeter part changes the y-coordinate direction to connect in the same direction in 2D space. Some of the U shapes are constituents of the C shape group. When a perimeter part satisfies the U and C shape conditions together, the U shape condition is given priority; we use that first. A point that is within the point set of a C shape voxel set related to implicit function theorem is selected in the middle of voxels that are vertical. The last classifying step is to select points related to the mathematical concept of the line integral. The line integral is understood as a linear connection of the points on the curve. For each ∆x that continues with the same y value in the perimeter voxel set, we select the mid-point – an 8-connected voxel set in 2D (26-connected voxels in 3D space). Each line integral part starts and ends with either a U, T, L, or C shape. When our understanding of the boundary involves the implicit function theorem, the boundary point uses a point in the C shape. When our understanding of the boundary involves the non-differential part, then the boundary point uses a point in any of the U, T, or L shapes.

\section{Conclusions}

\subsection{Throat-Finding Algorithms}

To find accurate throat, we calculate all possible areas with five algorithms. The results can classify the throat types. In natural samples, non-simply connected type is rare. We test our new algorithms with a hypothetical and then with real high porosity samples of more than 20$\%$. When two cylinders cross each other with a small cross area, the throat is saddle-shaped (Figure 11). Our algorithms suggest two possible throat candidates: a planar throat and a non-planar throat (Figure 11 A, B, and C). The throat we analyze is the smallest one (Figure 11 D), the actual. The second test sample is sphere pack with 9*9*9 spheres with a rectangular bounday. My algorithm can find all throats (Figure 12). Also, we test 8 of the S3 samples and 4 sections of the Handford wet inlet t241 samples. The results for these twelve tests are shown in Table 1. For cubic samples (512*512*512 $voxel^{3}$), the total CPU time consumed for the throat finding algorithms and probability density function of throat area, pore volume, and coordination number is 4 to 10 hours. Our new algorithms include a crossed-throat algorithm [4]. The Pdf of throat area, pore volume, and coordination number is shown in Figure 13.

\begin{center}
\epsfig{figure=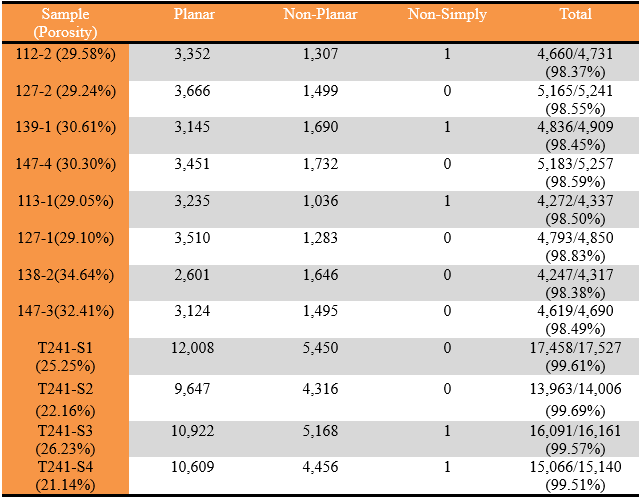,height=4.6in,width=6.0in} \ \ \ \ \ \ \  \small{Table 1 : Throat type and success ratio of the new algorithm}
\end{center}


\subsection{Point Set Algorithm}
To select a point set in 3D space, we project the real planar perimeter voxel set (in 3D space) onto the axial plane (2D) consisting of two coordinates, the lowest two absolute values among three components of the normal vector of the plane. To test our new algorithm, we apply our algorithm to lines with different slopes and circles with different radii. For the linear boundary, we draw the linear boundary $y=tan⁡(k^{o})x$, where $k$=1,2,⋯,89, and the domain is from 0 to [50,000 $cos⁡(k^{o})$ ], where [x] is a maximum integer less than or equal to x. We make digitized images of void and grain voxel areas. The grain voxels are selected when we have more than half of the voxel area below the linear boundary. The mid-point algorithm’s maximum relative error for the linear boundary is 8.23$\%$, and the new algorithm’s relative error is 1.29$\%$ (Figure 14 C). The circular boundary has two perimeter parts that relate to the C shape, and the boundary doesn’t have any U, T, or L shapes. We can apply two line integrals which indicate the upper boundary and the lower boundary. The mid-point algorithm’s relative error oscillates around 5.5$\%$, but the new algorithm has less than a 1$\%$ relative error (Figure 14 D). For non-planar throat type, triangular throat area incluiding a medial axis voxel can over-calculate its area (Figure 15 The orange ellipse area is my interested area. B, C, and D are the same image with triangular throat calculation including a medial axis voxel. E is my preferred solution.)

\newpage
\Large{\bf{Figures}}

\begin{center}
\epsfig{figure=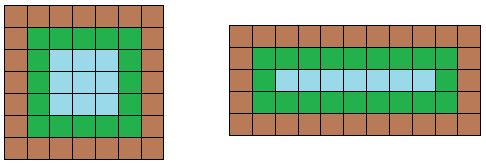,height=1.3in,width=4.7in}
\end{center}
\small{Figure 1 :  It shows that the short path does not always indicate the small area. The area (light blue color) of a throat in the right panel is smaller than the area in the left panel but the distance of perimeter of boundary voxel (green color) is opposite. (Brown color : grain voxel, green color : perimeter and boundary grain voxel and light blue color : throat barrier voxels)} 

\begin{center}
\epsfig{figure=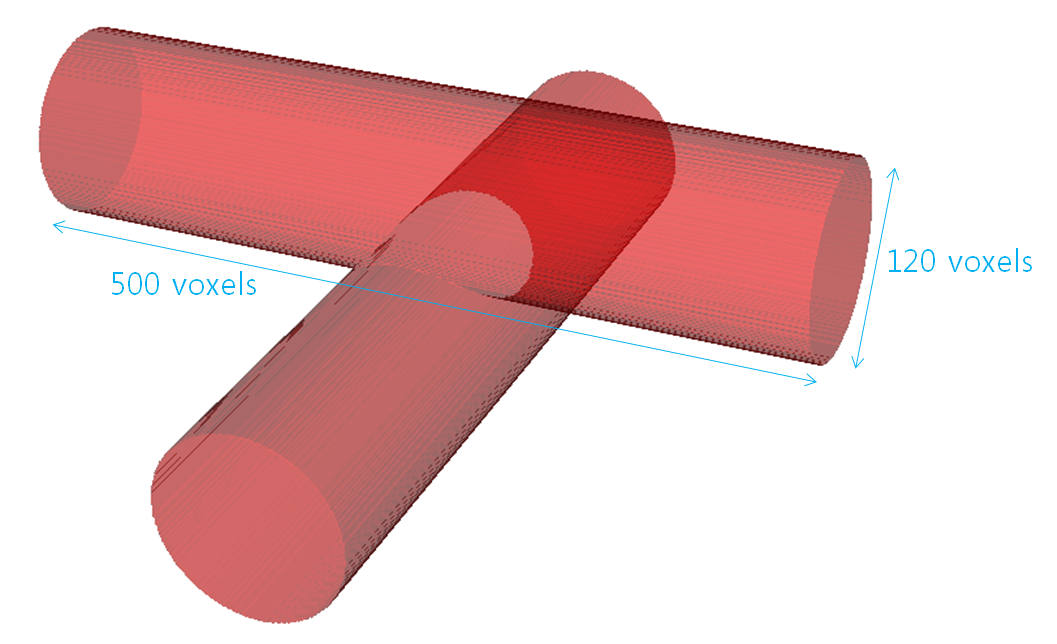,height=1.8in,width=2.8in}
\epsfig{figure=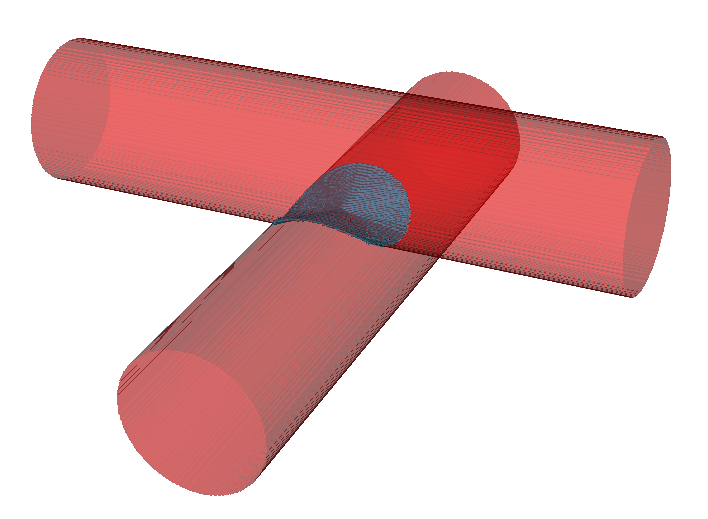,height=1.8in,width=2.8in}
\end{center}
\small{Figure 2 : Example of Two cross cylinders: left figure is an example of non-planer throat type and the right figure shows the real throat (blue color) of left fiugre.} 

\begin{center}
\epsfig{figure=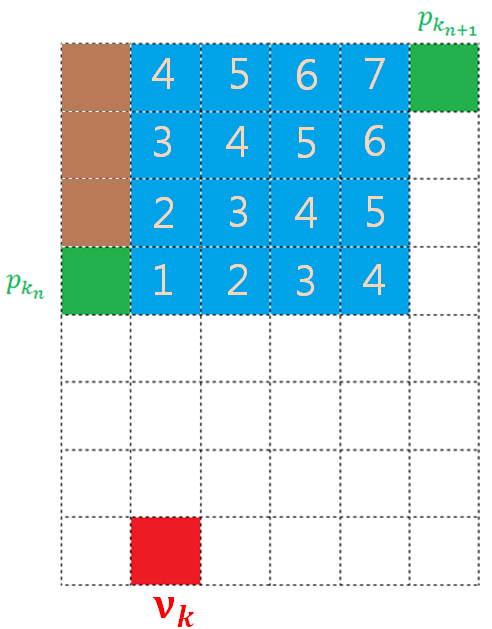,height=1.8in,width=2.1in}
\epsfig{figure=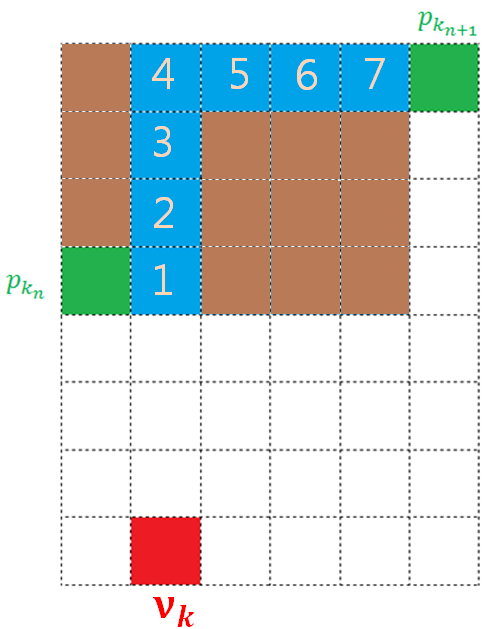,height=1.8in,width=2.1in}
\epsfig{figure=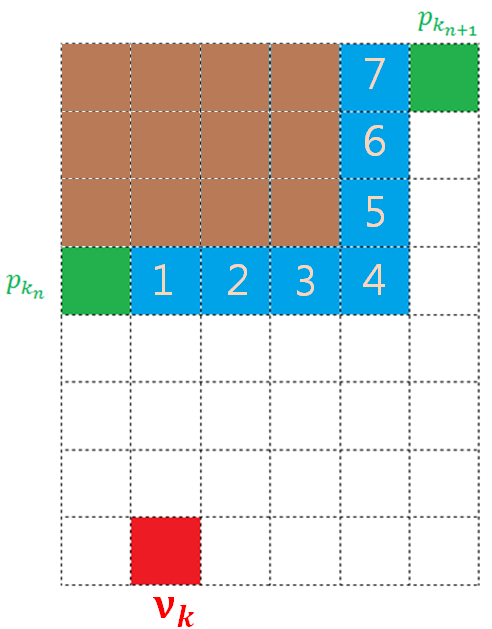,height=1.8in,width=2.1in}
\end{center}
\small{Figure 3 : Test example for Dijkstra's algorithm and possible paths. Blue voxels are possible path grain voxels. Green color indicates the initial and terminal grain voxels. Left panel shows all possible paths using Dijkstra's algorithms and numbers indicate step number. Middle panel shows an example of errant approach using Dijkstra's algorithm with smallest tirangular area and Right panel illustrates real boudary grain path using Dijkstra's algorithm with shortest distance for each step voxel. }

\begin{center}
\begin{tabular}{ll}
{\bf A}&{\bf B}\\
\epsfig{file=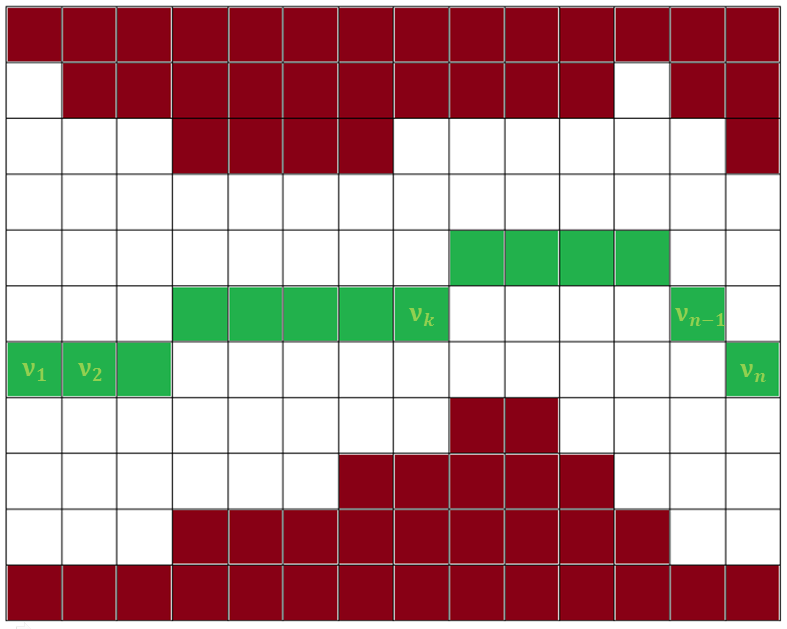, height=2.0in, width=3.1in}&
\epsfig{file=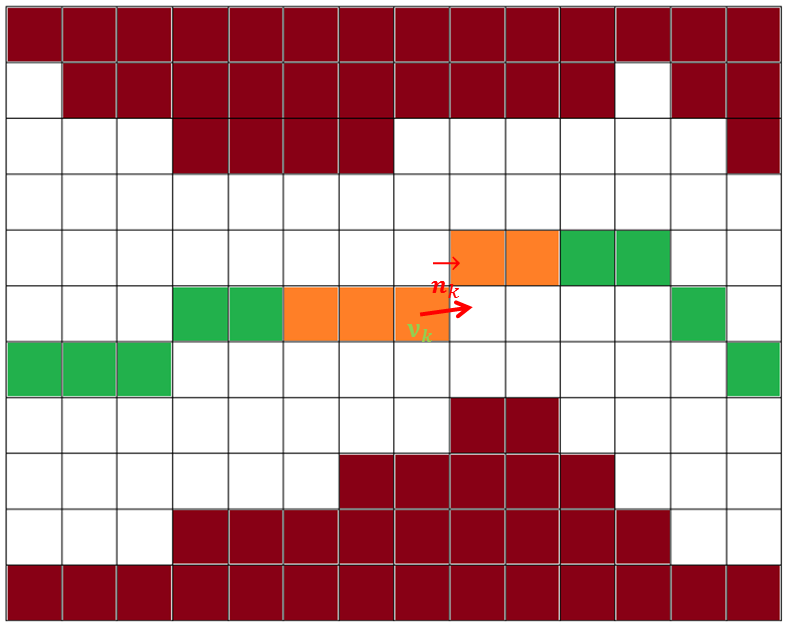, height=2.0in, width=3.1in}\\
{\bf C}&{\bf D}\\
\epsfig{file=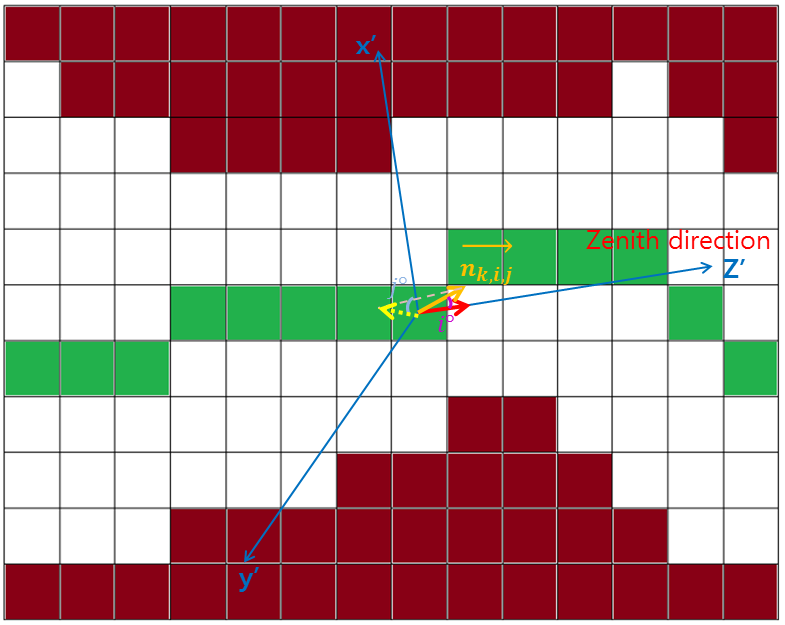, height=2.0in, width=3.1in}
&\epsfig{file=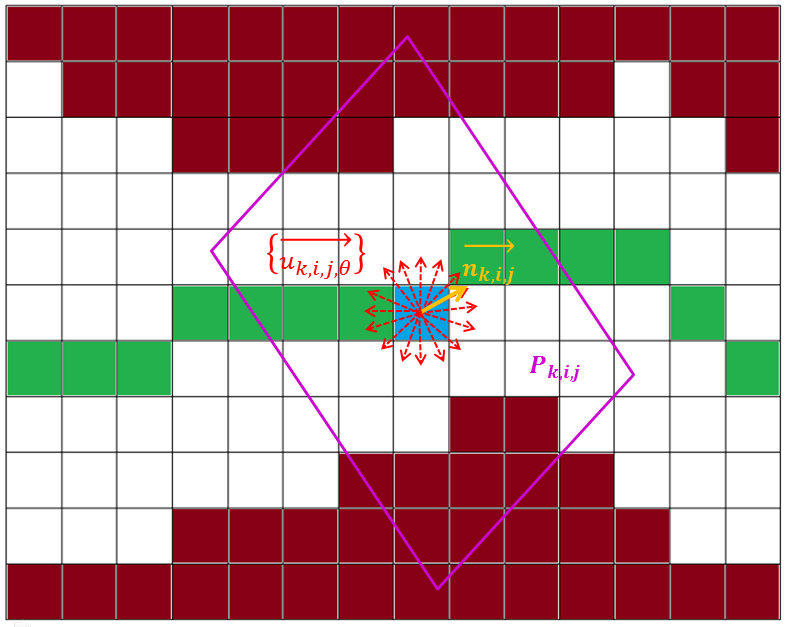, height=2.0in,
width=3.1in}\\
{\bf E}&{\bf F}\\
\epsfig{file=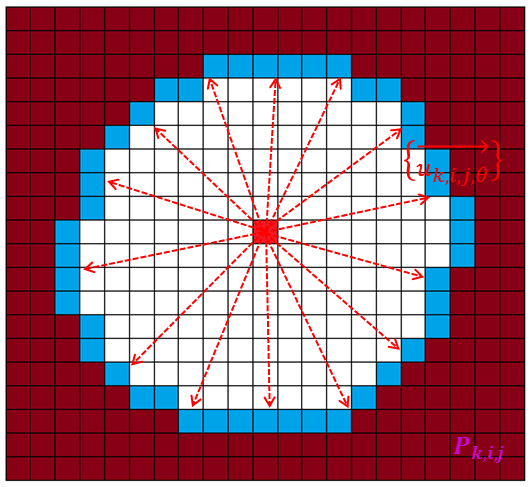, height=2.0in, width=3.1in}
&\epsfig{file=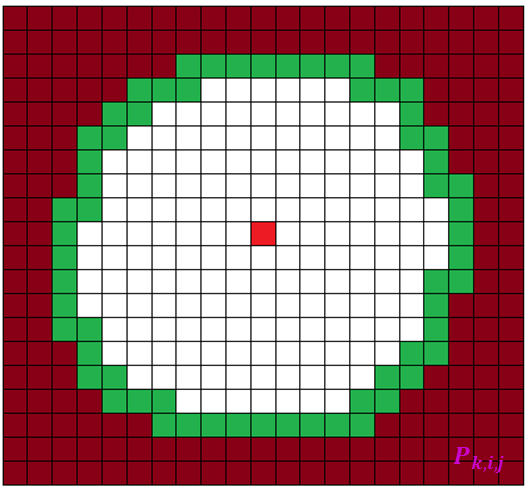, height=2.0in,
width=3.1in}\\
\end{tabular}
\end{center}
\small{Figure 4 : The first algorithm processing diagram. A : vertical subsection slide of 3D image (dark red, green and white colors indicate the grain, the medial axis, and void space, respectively). B : finding the tangent vector $\protect\overrightarrow{n_{k}}$ at the center of $\nu_{k}$ using five consecutive voxels. C : finding  unit directional vector $\protect\overrightarrow{n_{k,i,j}}$. This vector is an unit directional vector in spherical coordinate system whose zenith directions is in the same direction of  $\protect\overrightarrow{n_{k}}$. D : constructing a plane, $P_{k,i,j}$ with both a point (the center of $\nu_{k}$) and a normal vector $\protect\overrightarrow{n_{k,i,j}}$ and an unit directional vector set \{$\protect\overrightarrow{u_{k,i,j,\theta}}$\} on the plane $P_{k,i,j}$.   E :  finding the boundary grain voxel set using the unit directional vector set \{$\protect\overrightarrow{u_{k,i,j,\theta}}$\}. F : changing the connectivity from 26 to 6 connectivity of the founded boundary grain voxel set from E.  } 

\begin{center}
\epsfig{figure=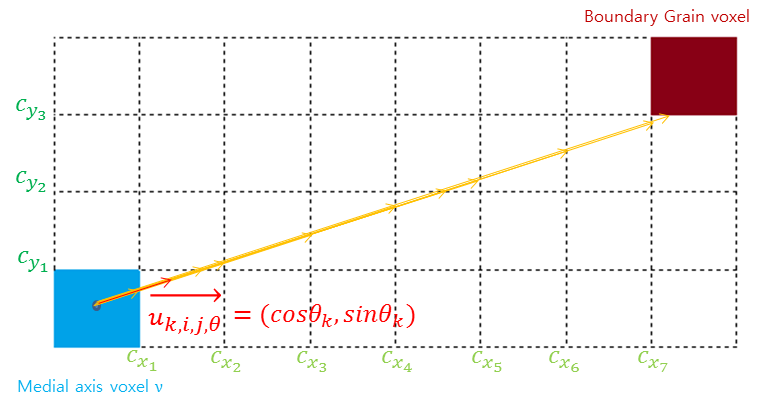,height=3.2in,width=6.2in}
\end{center}
\small{Figure 5 : A unit ray spread in the digitized image. Yellow colored vector is in the direction of the unit vector $\protect\overrightarrow{u_{k,i,j,\theta}}$ and touches the boundary grain voxel.}

\begin{center}
\begin{tabular}{ll}
{\bf A}&{\bf B}\\
\epsfig{file=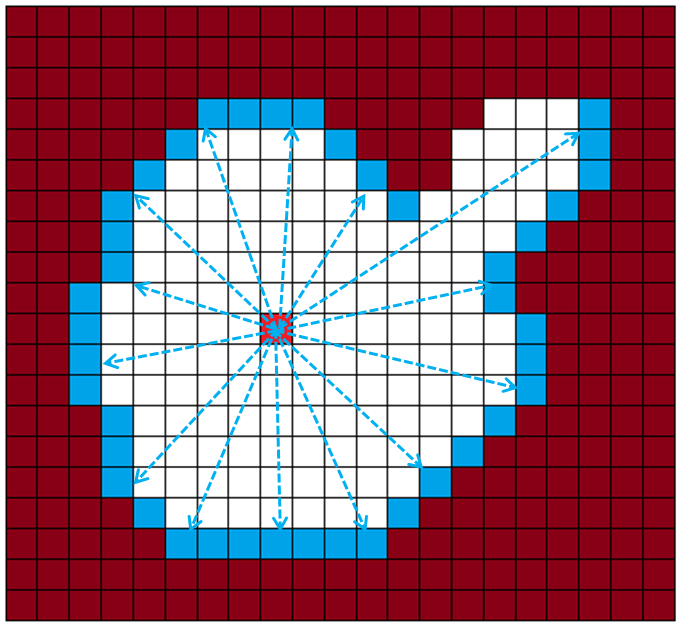, height=2.2in, width=3.3in}&
\epsfig{file=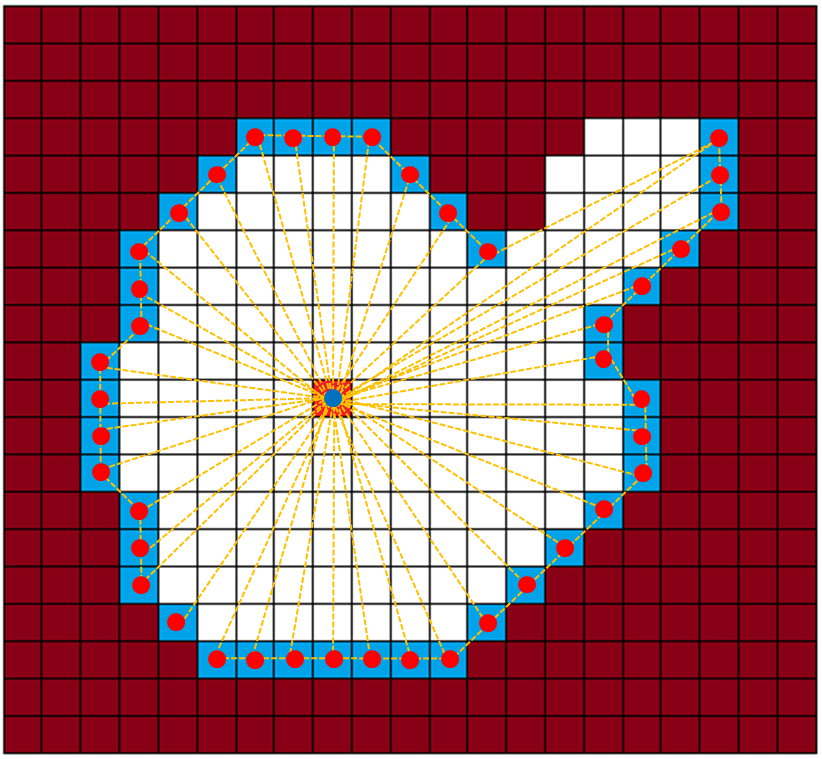, height=2.2in, width=3.3in}\\
{\bf C}&{\bf D}\\
\epsfig{file=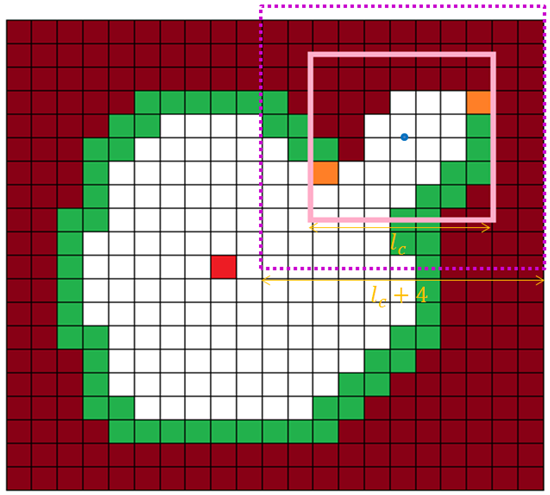, height=2.2in, width=3.3in}
&\epsfig{file=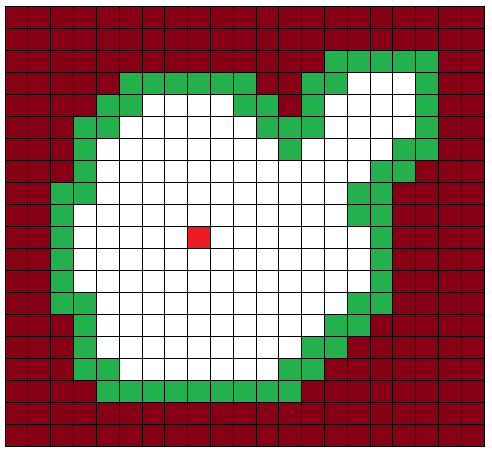, height=2.2in,
width=3.3in}\\
\end{tabular}
\end{center}
\small{Figure 6 : The second algorithm processing diagram. A : $\protect\overrightarrow{u_{k,i,j,\theta}}$ detects the visible boundary grain voxels (blue colored voxels). B : Finding triangular area using both two centers of boudary grain voxels and the center of medial voxel (red colored voxel in the middle). C : The minimum (pink rectangle) cube encompasses two non-connected (orange color) voxels. The side length of the cube is $l_{c}$. The interested region is between $l_{c}$+ 4 and $l_{c}$+8 to find accurate throats. D : Final boundary grain voxels using the second algorithm. } 

\newpage
\begin{center}
\begin{tabular}{ll}
{\bf A}&{\bf B}\\
\epsfig{file=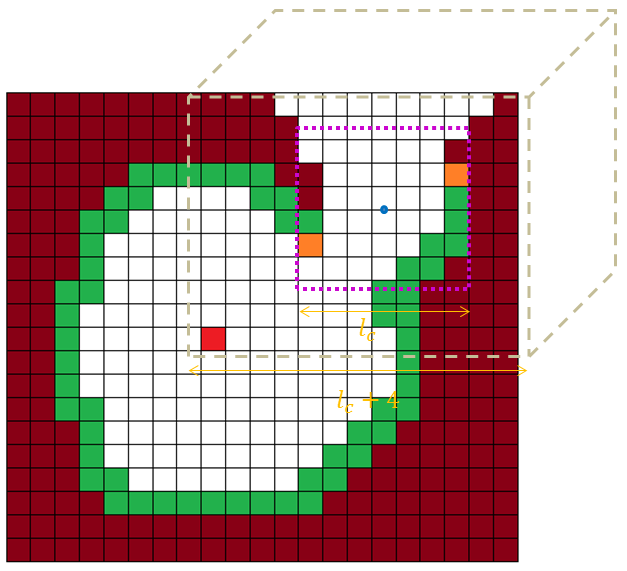, height=2.2in, width=3.3in}&
\epsfig{file=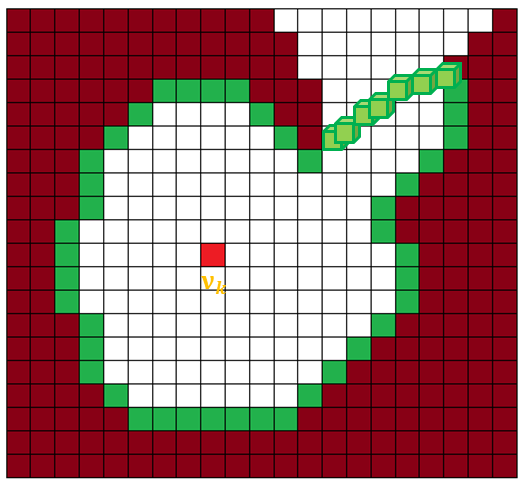, height=2.2in, width=3.3in}\\
\end{tabular}
\end{center}
\small{Figure 7 : Detecting the boundary grain path in space using Dijkstra's algorithm. A : Dijkstra's algorithm is applied to interested region cube in space. B : detected boundary grain path in space. } 

\newpage
\begin{center}
\begin{tabular}{ll}
{\bf A}&{\bf B}\\
\epsfig{file=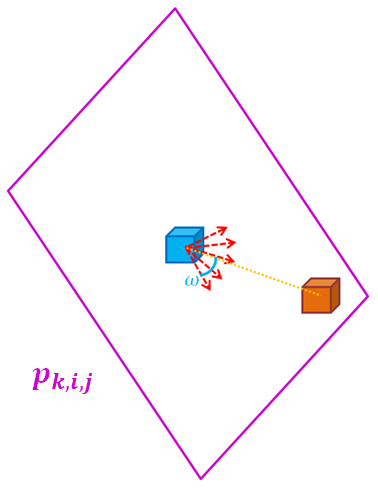, height=2.5in, width=3.3in}&
\epsfig{file=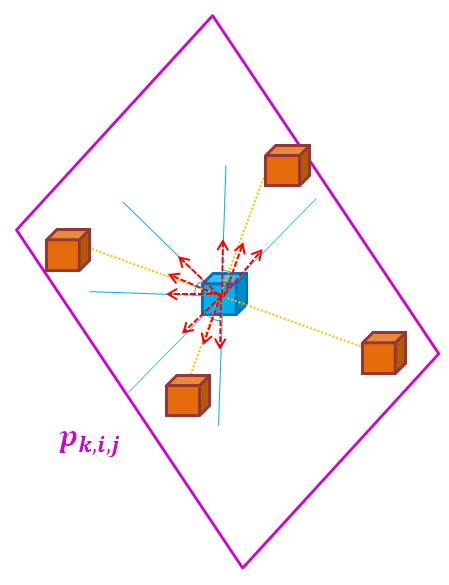, height=2.5in, width=3.3in}\\
{\bf C}&{\bf D}\\
\epsfig{file=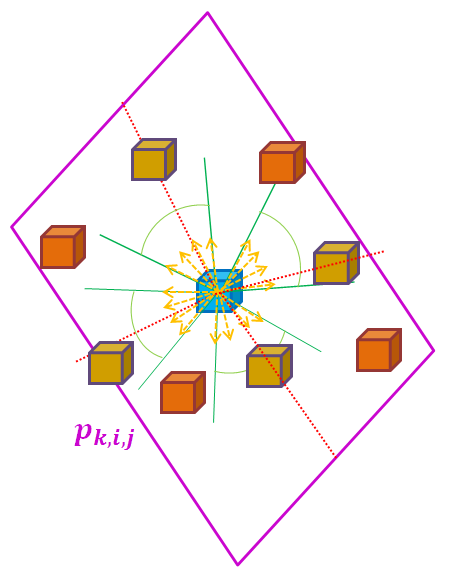, height=2.5in, width=3.3in}
&\epsfig{file=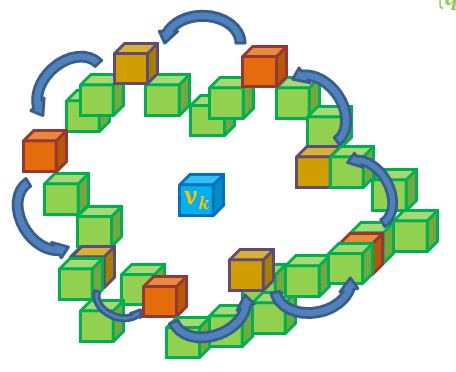, height=2.5in,
width=3.3in}\\
\end{tabular}
\end{center}
\small{Figure 8 : The processing of non-planar algorithm. A : detection of the nearest boundary grain voxel using $\protect\overrightarrow{u_{k,i,j,\theta}}$ where $\theta$ is between $0^{o}$ and $90^{o}$. B : Detecting another nearest boundary grain voxel with different $\theta$. C : make four verticle wedge (green ray) and dectect the boudary grain voxels not on the plane $P_{k,i,j}$ .  D : Detected the boundary grain voxels in space. }

\begin{center}
\begin{tabular}{ll}
{\bf A}&{\bf B}\\
\epsfig{file=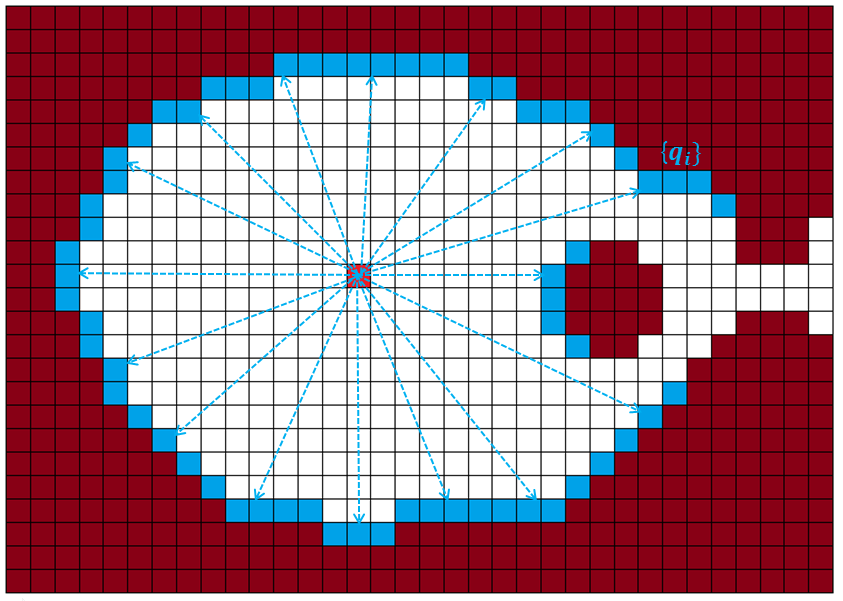, height=1.7in, width=2.7in}&
\epsfig{file=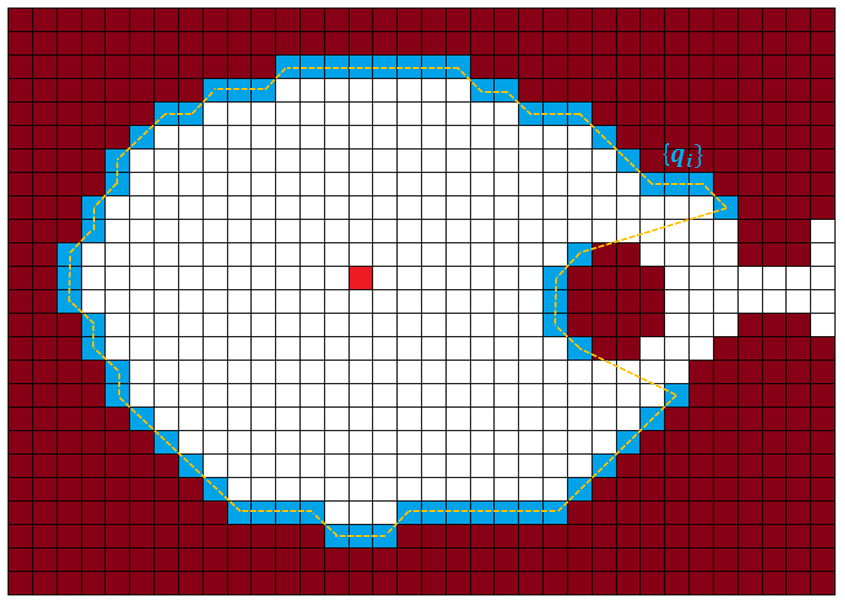, height=1.7in, width=2.7in}\\
{\bf C}&{\bf D}\\
\epsfig{file=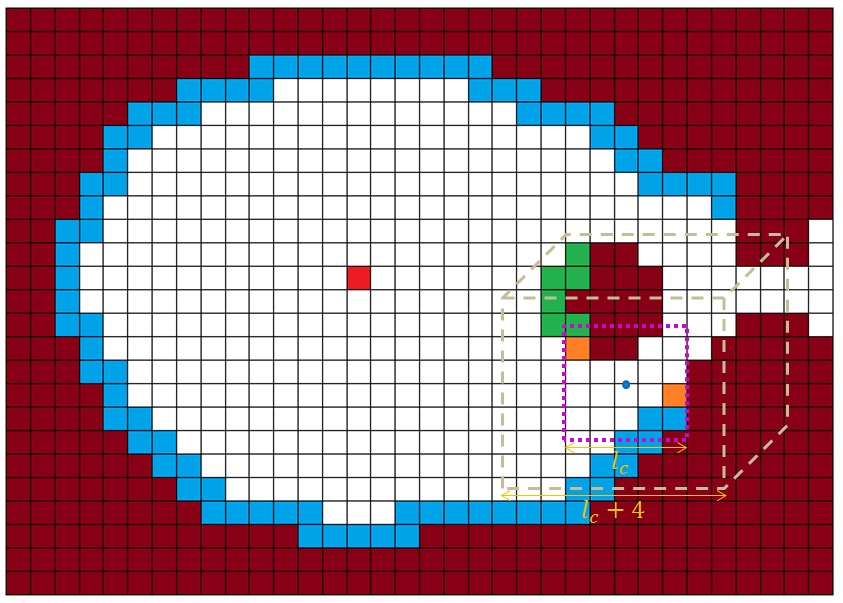, height=1.7in, width=2.7in}
&\epsfig{file=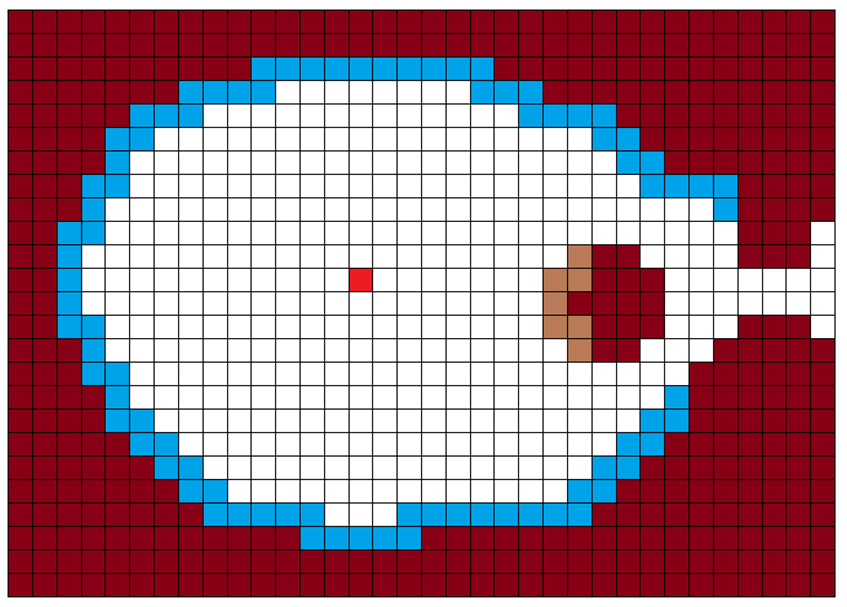, height=1.7in,
width=2.7in}\\
{\bf E}&{\bf F}\\
\epsfig{file=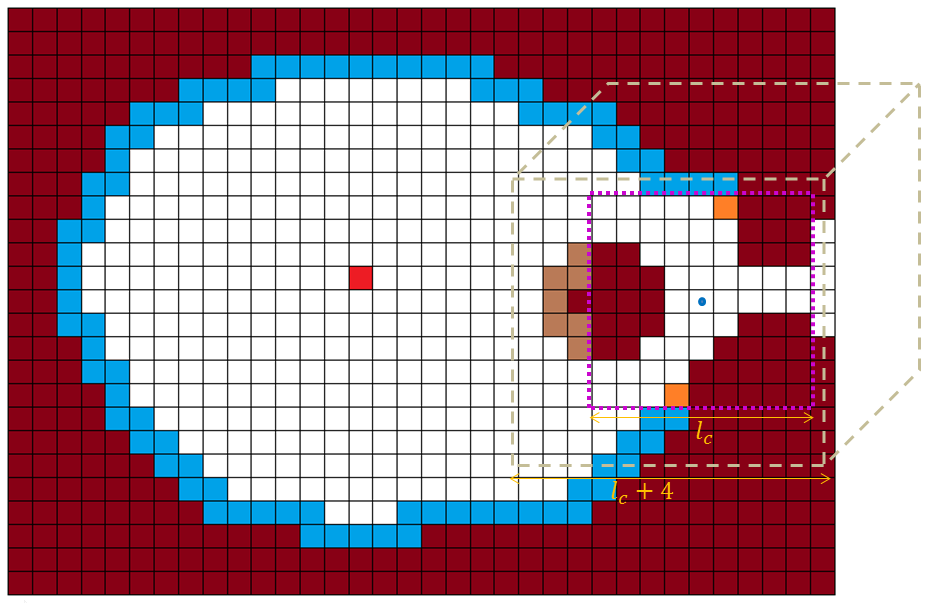, height=1.8in, width=2.7in}&
\epsfig{file=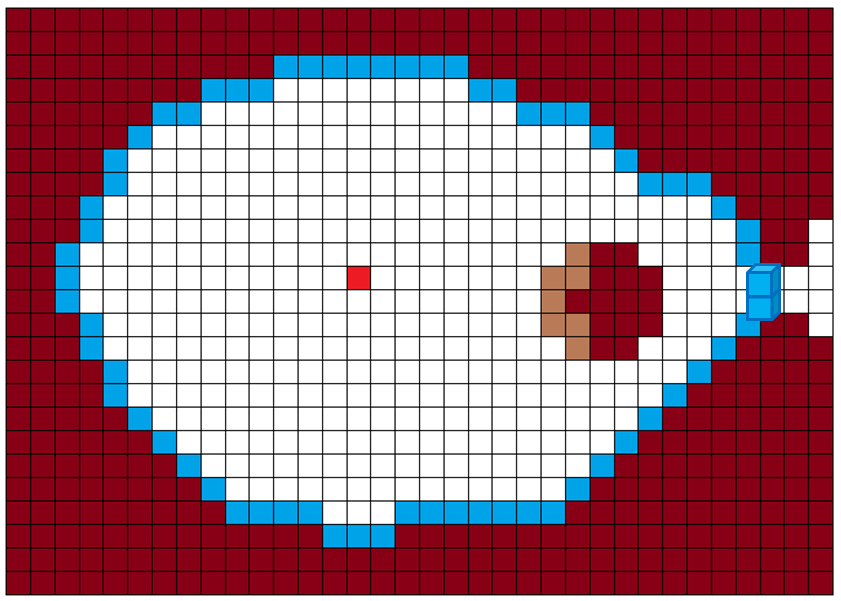, height=1.8in, width=2.7in}\\
{\bf G}&{\bf H}\\
\epsfig{file=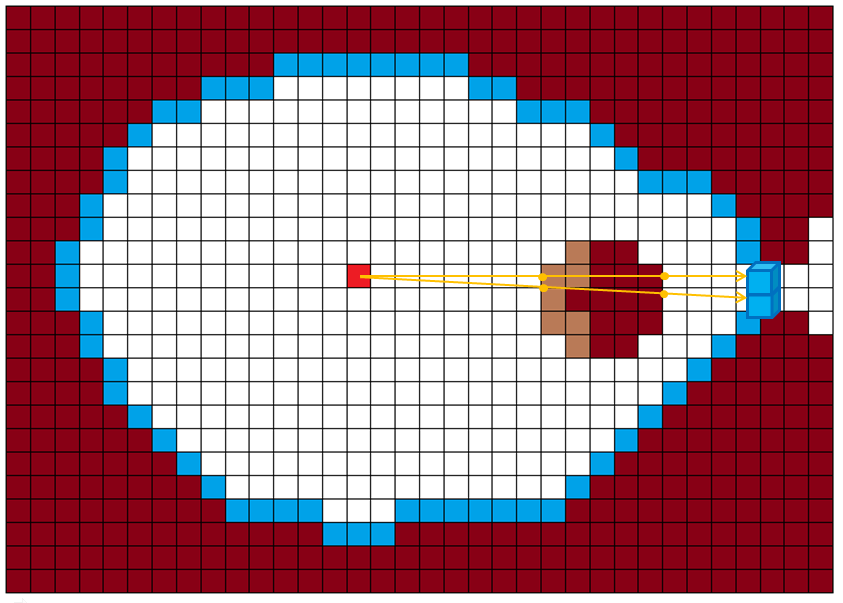, height=1.8in, width=2.7in}
&\epsfig{file=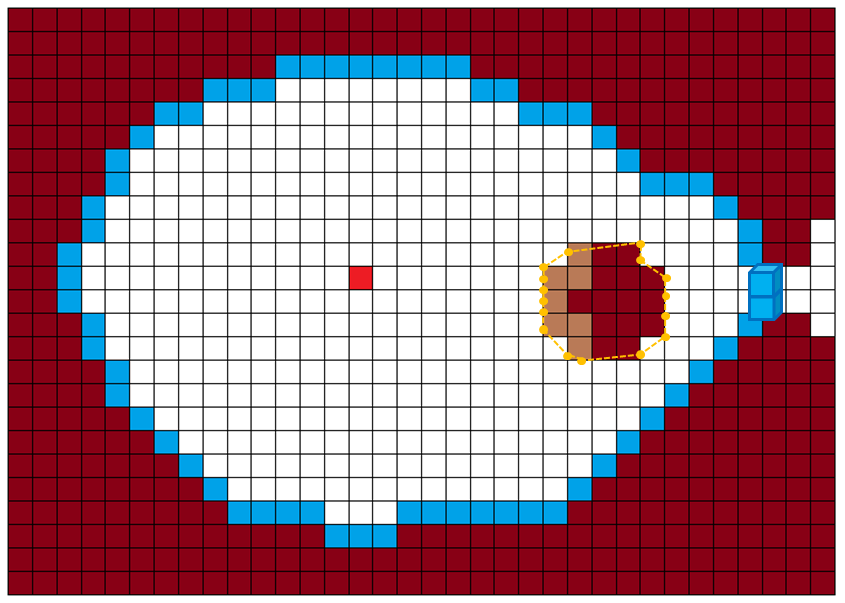, height=1.8in,
width=2.7in}\\
\end{tabular}
\end{center}
\small{Figure 9 : Non planar algorithm processing diagram. A: visible voxel sets (blue) from the medial axis on a plane. B : calculating the sum of triangular area using both the visible voxel sets and a medial voxel ($\nu_{k}$). C : connect to disjoint consecutive voxel using Dijkstra's algorithm. D : nixed voxel set (brown voxels). E : New Region of interest. F : Outer 26 connected perimeter voxel set. G : Find 4 points on the boundary of inner grain voxels. H : The inner grain boundary and the outer perimeter voxel set. }

\begin{center}
\begin{tabular}{ll}
{\bf A}\\
\epsfig{file=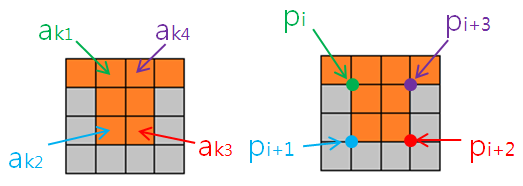, height=2.1in, width=5.2in}\\
{\bf B}\\
\epsfig{file=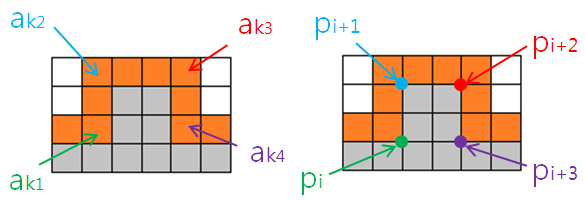, height=2.1in, width=5.1in}\\
{\bf C}\\
\epsfig{file=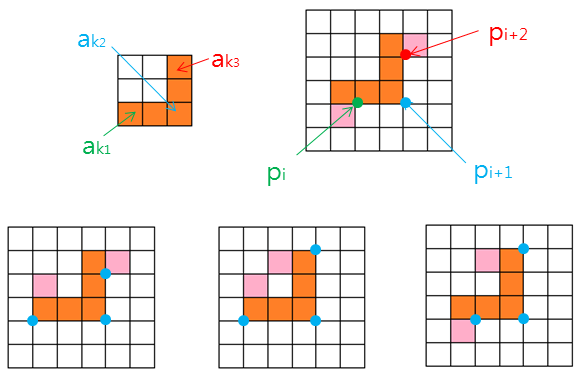, height=3.3in, width=5.0in}\\
\end{tabular}
\end{center}
\small{Figure 10 : The method how to find points with respect to non-differential section. A : Perimeter voxels and its points set in the T shape. B : Perimeter voxels and its points set in the U shape. C : Perimeter voxels and its points set in the L shape.} 

\begin{center}
\begin{tabular}{ll}
{\bf A}&{\bf B}\\
\epsfig{file=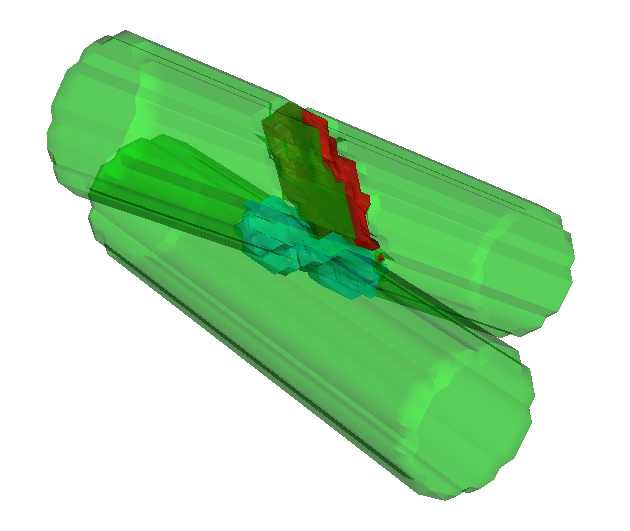, height=2.5in, width=3.3in}&
\epsfig{file=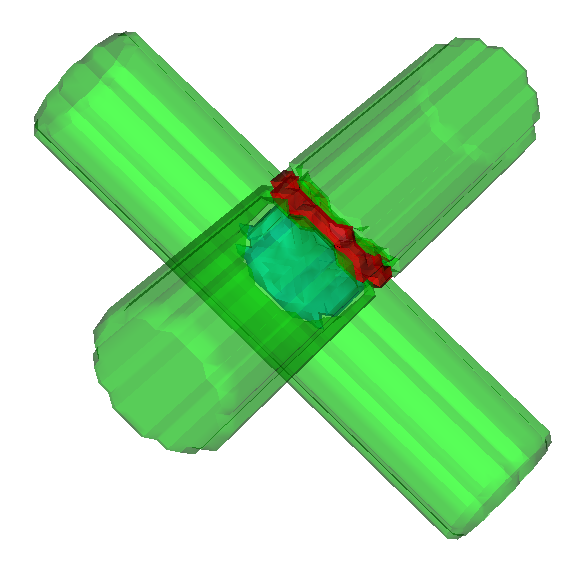, height=2.5in, width=3.6in}\\
{\bf C}&{\bf D}\\
\epsfig{file=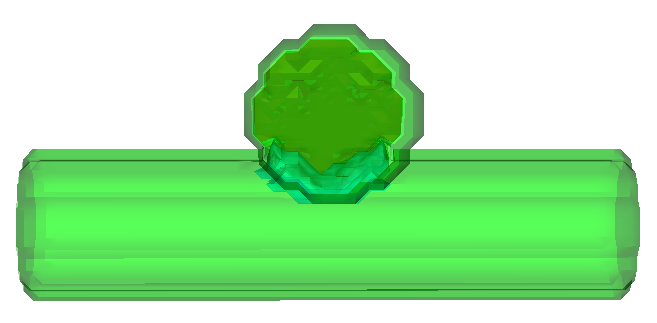, height=2.1in, width=3.4in}
&\epsfig{file=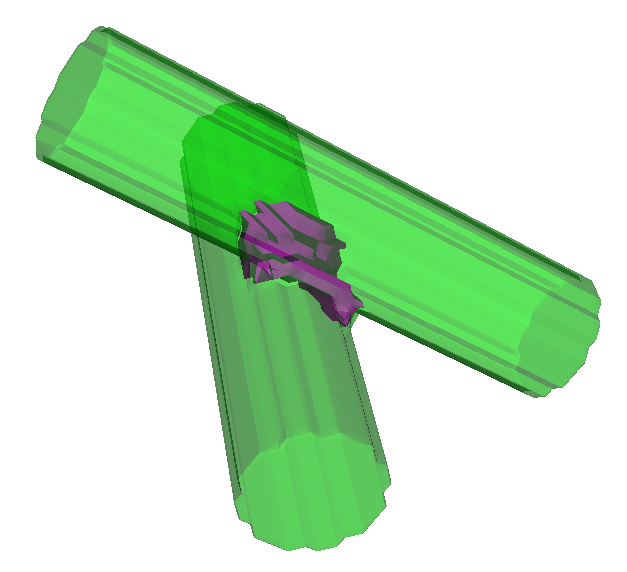, height=2.5in,
width=3.3in}\\
\end{tabular}
\end{center}
\small{Figure 11: Planar and non-planar throat types. Presented algorithms in this paper detect both planar and non-planar throat (red and light blue color in A and B), then I choose the ideal throat by taking the minimum of both detected planar and non-planar throats.  Red color :  planar throat barrier voxels. light blue color : non-planar throat barrier voxel set, purple color : final throat barrier voxel set }

\begin{center}
\begin{tabular}{ll}
{\bf A}&{\bf B}\\
\epsfig{file=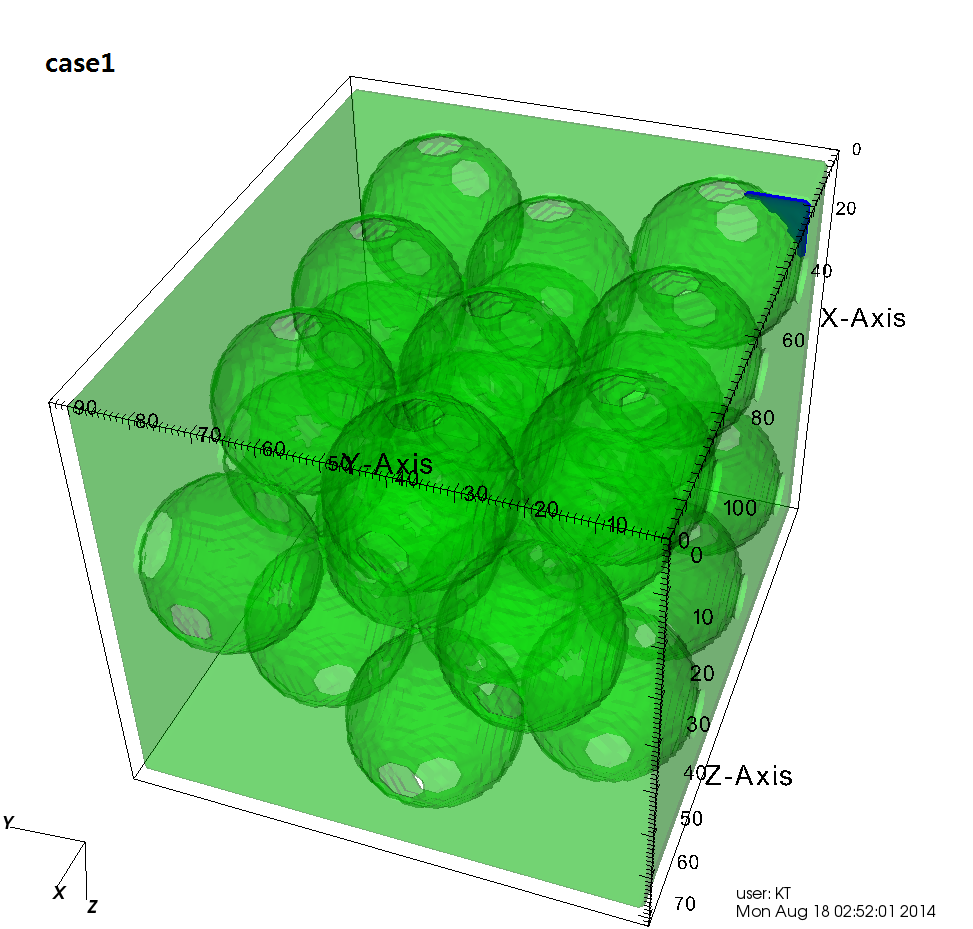, height=2.0in, width=3.3in}&
\epsfig{file=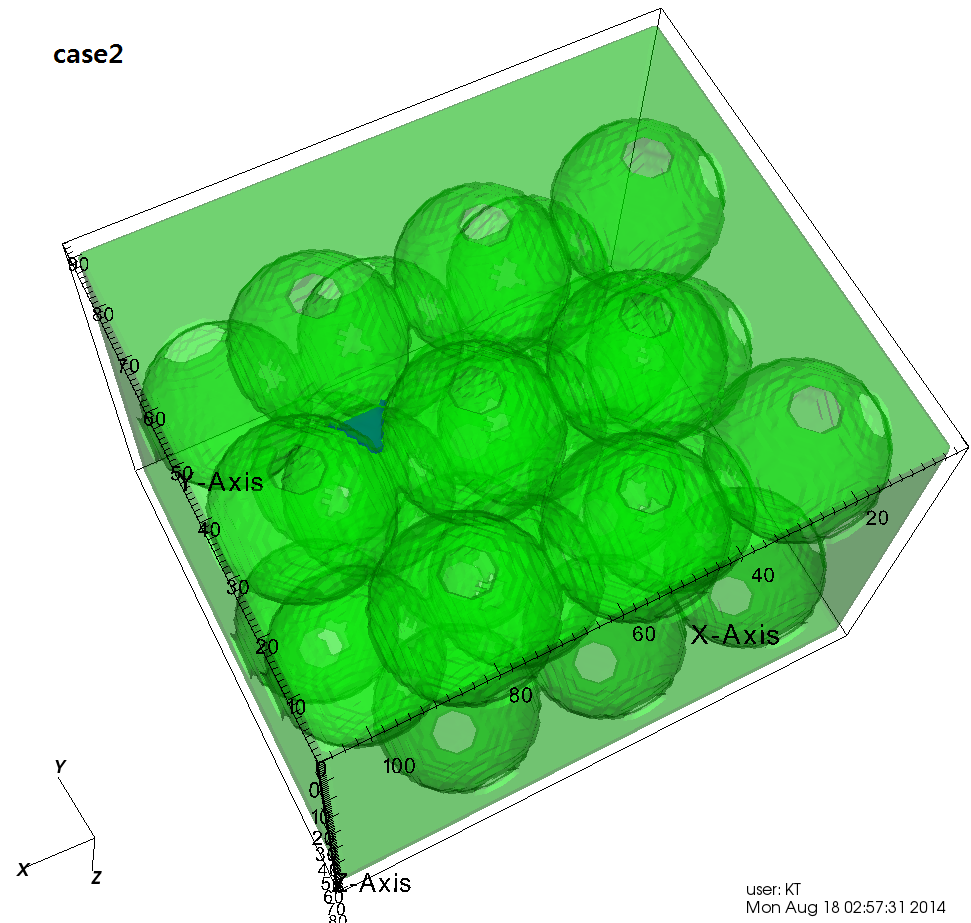, height=2.0in, width=3.3in}\\
{\bf C}&{\bf D}\\
\epsfig{file=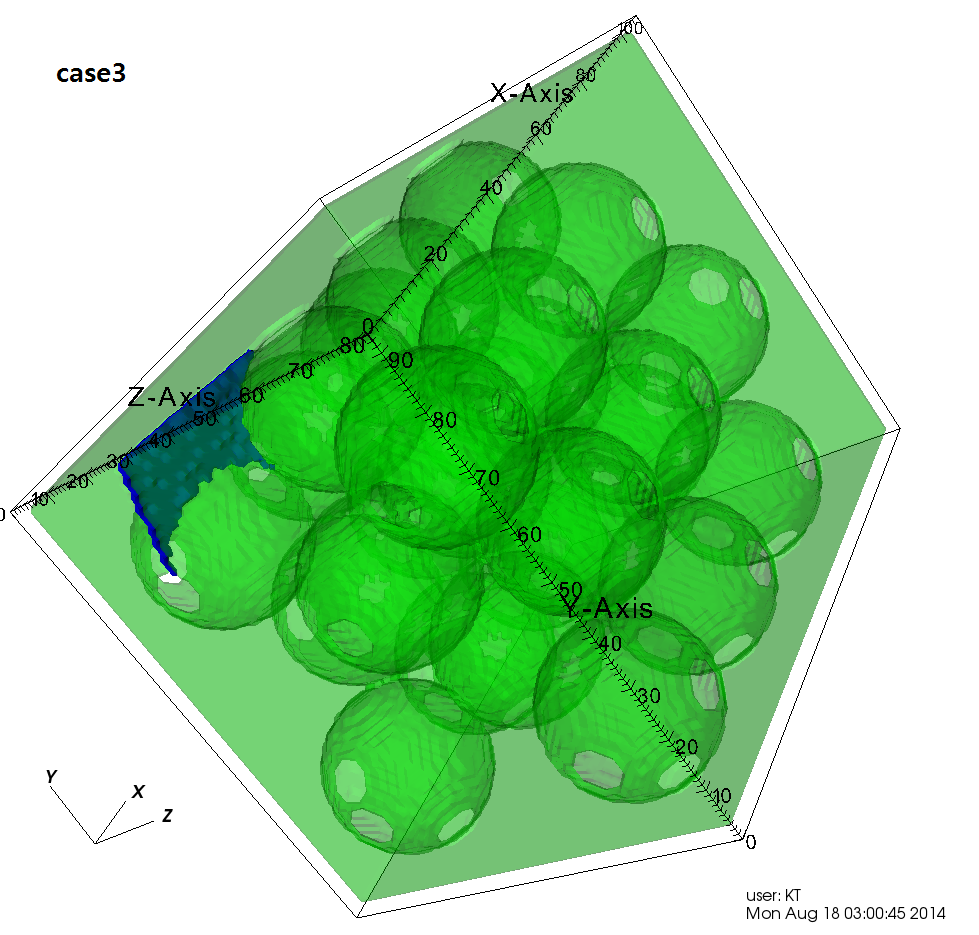, height=2.0in, width=3.3in}
&\epsfig{file=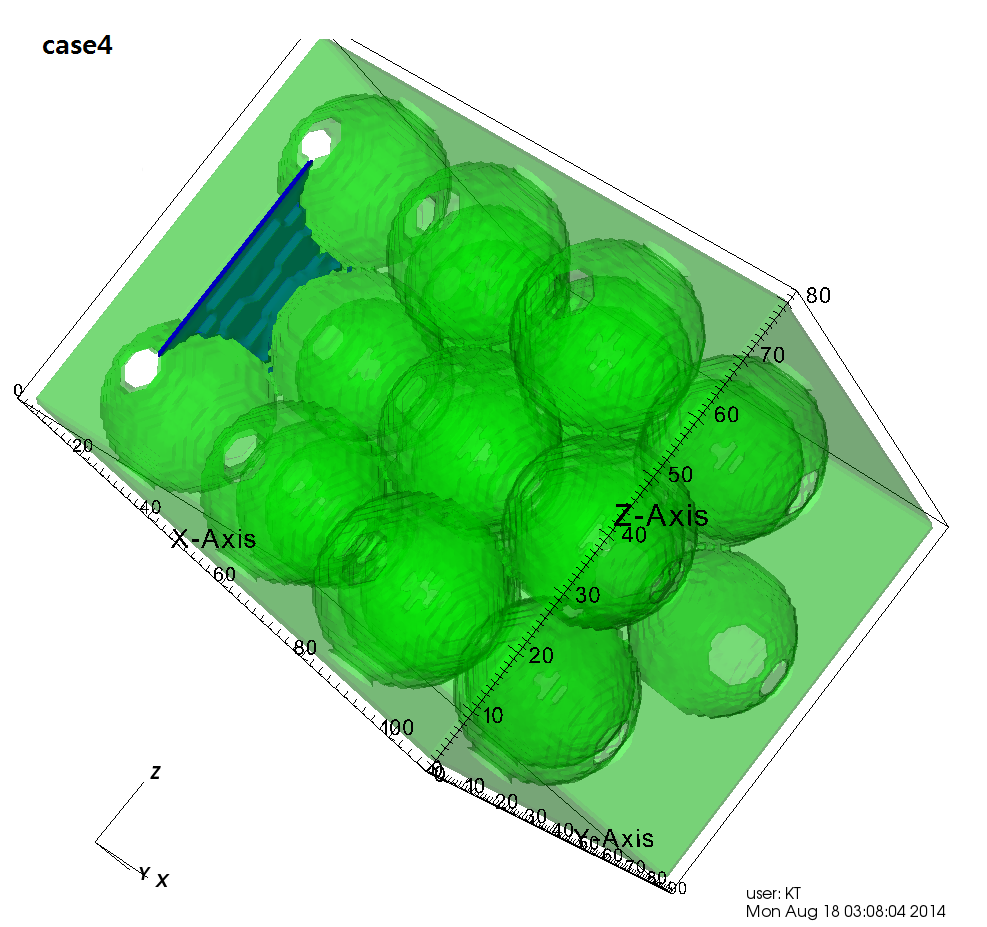, height=2.0in,width=3.3in}\\
{\bf E}\\
\epsfig{file=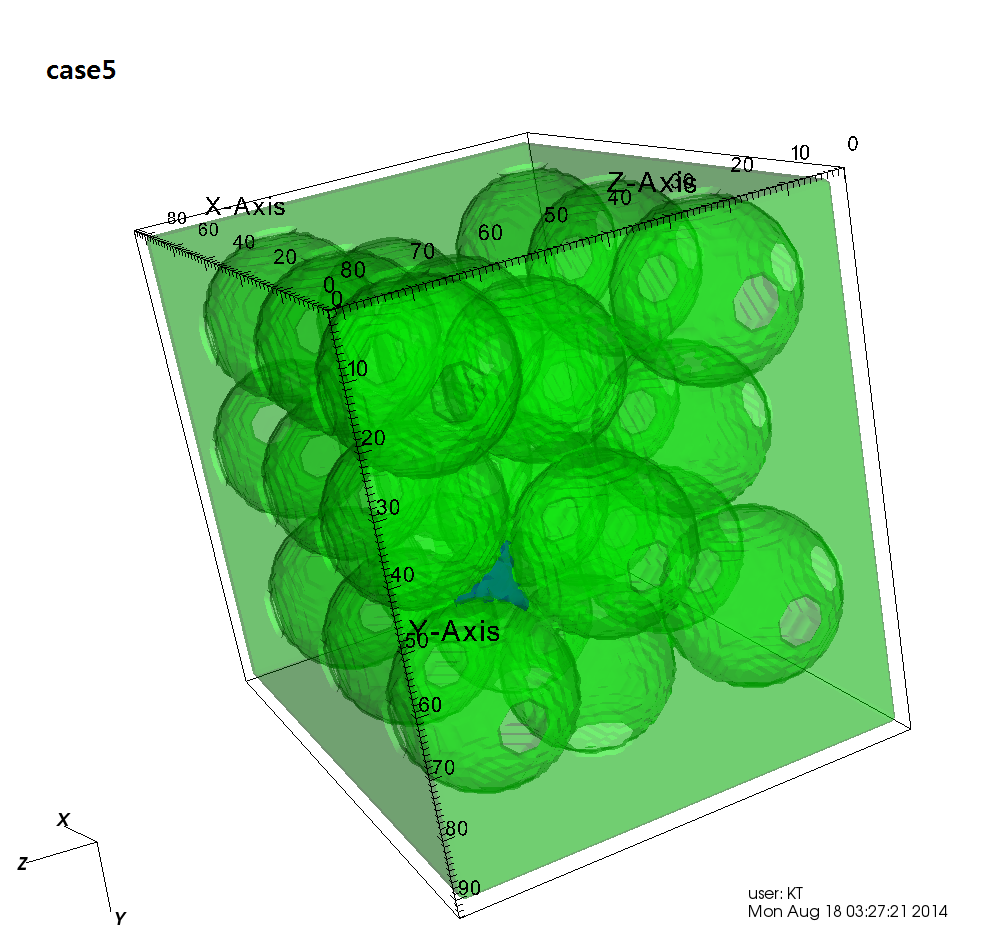, height=2.0in, width=3.3in}
\\
\end{tabular}
\end{center} 
\small{Figure 12 : Sphere pack sample with 5 representative throats. Blue color area indicate one of five representative throats. This sample has 3 layer spheres and each layer has 9 spheres.

\begin{center}
\begin{tabular}{ll}
{\bf A}&{\bf B}\\
\epsfig{file=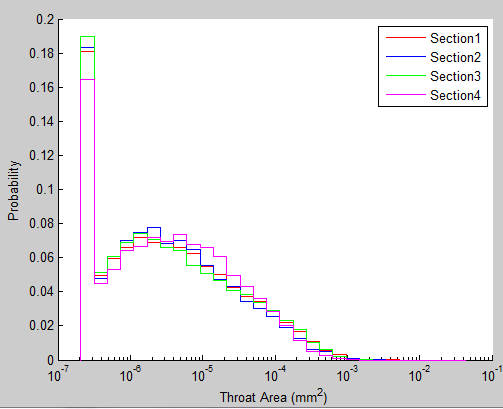, height=2.0in, width=3.1in}&
\epsfig{file=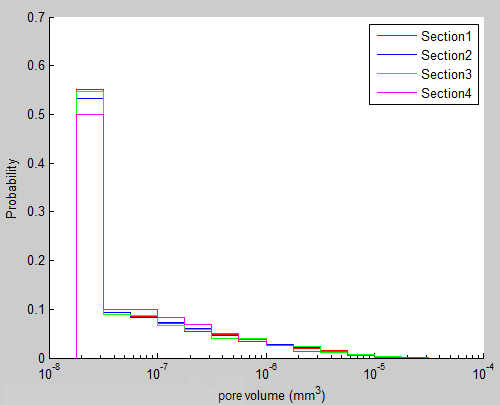, height=2.0in, width=3.1in}\\
{\bf C}\\
\epsfig{file=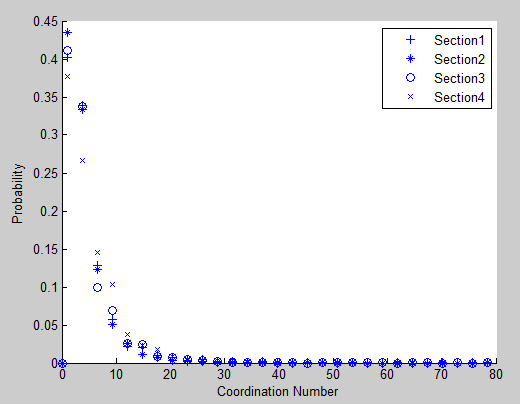, height=2.0in, width=3.1in}\\
\end{tabular}
\end{center}
\small{Figure 13 : Throat area (A), Pore volume distribution (B), and Coordination number (C) computed from the XCMT image of the Handford wet inlet t241 samples using 5 algorithms  }

\begin{center}
\begin{tabular}{ll}
{\bf A}&{\bf B}\\
\epsfig{file=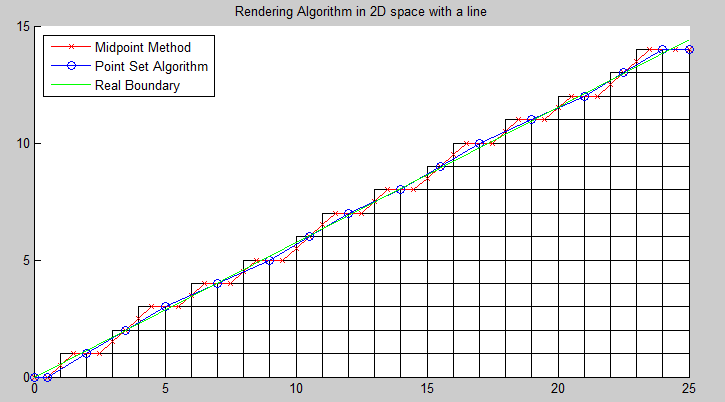, height=2.5in, width=3.3in}&
\epsfig{file=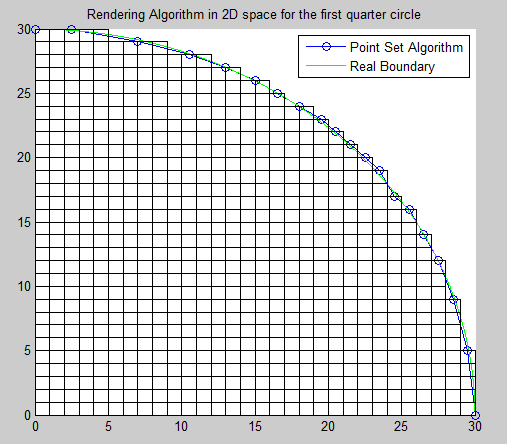, height=2.5in, width=3.3in}\\
{\bf C}&{\bf D}\\
\epsfig{file=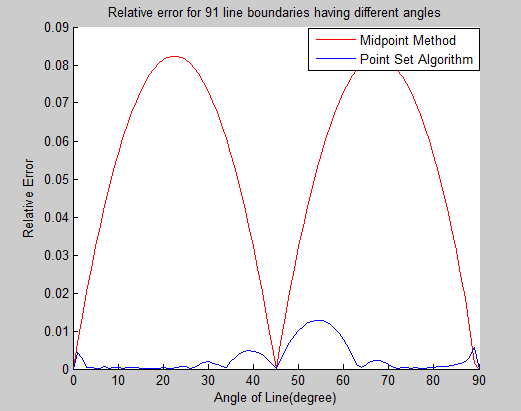, height=2.5in, width=3.3in}
&\epsfig{file=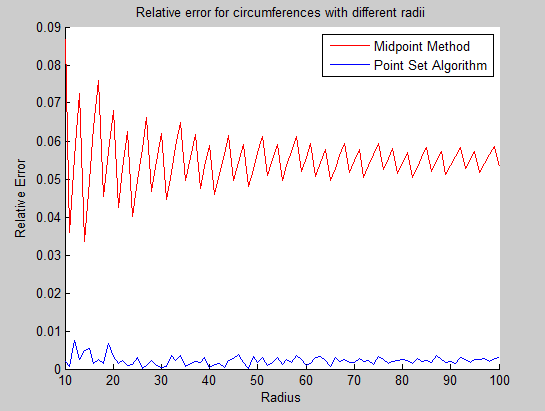, height=2.5in,
width=3.3in}\\
\end{tabular}
\end{center}
\small{Figure 14 : A,B : Example of calculating the exact linear distance (A) and circumferences (B) using the point set algorithm.  C,D : Relative error in the digitized image of the real boundary (a) line with different angel. (b) circle with different radii.}

\begin{center}
\begin{tabular}{ll}
{\bf A}&{\bf B}\\
\epsfig{file=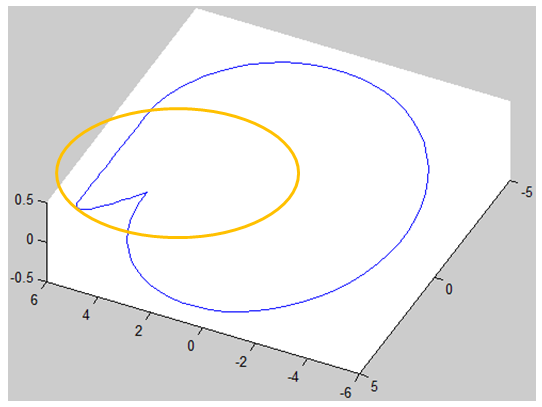, height=2.3in, width=3.3in}&
\epsfig{file=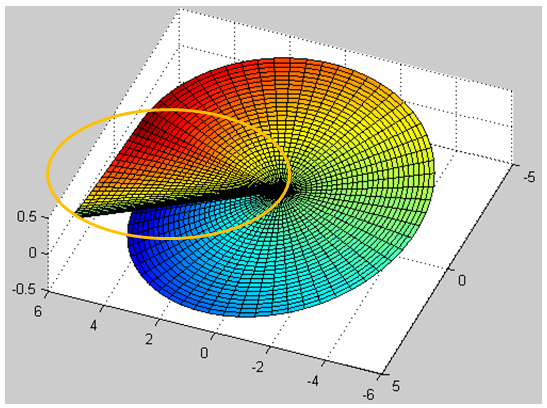, height=2.3in, width=3.3in}\\
{\bf C}&{\bf D}\\
\epsfig{file=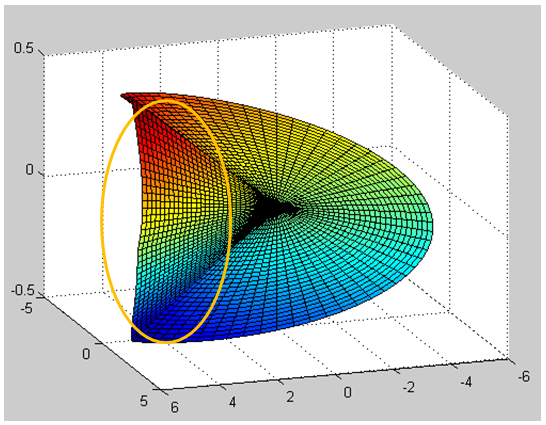, height=2.3in, width=3.3in}
&\epsfig{file=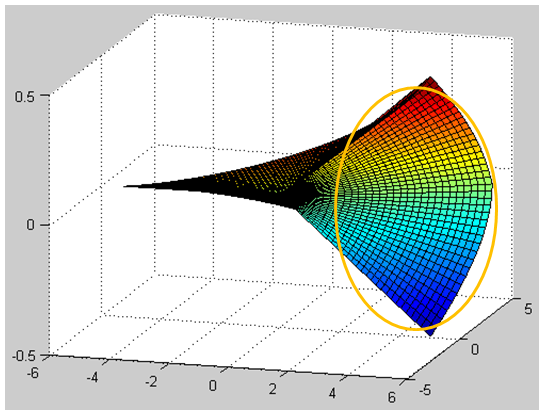, height=2.3in,
width=3.3in}\\
{\bf E}\\
\epsfig{file=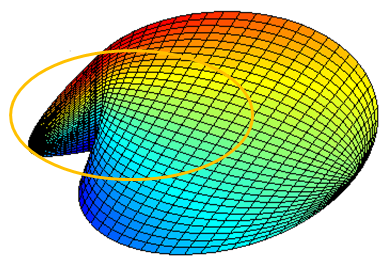, height=2.3in, width=3.3in}\\
\end{tabular}
\end{center}
\small{Figure 15 : Counter example to find throat area using the sum of triangular area (in 3DMA thorat area calculation algorithm). Existing algorithm can not calculate the exact throat area of this surface. A : blue colored curve shows a perimeter of the surface. 3DMA throat algorithm cannot calculate the exact throat in orange colored region. C, D : different point of view of B. E : desired throat surface.  }

\end{document}